\documentclass[pre,twocolumn,groupedaddress,showpacs,amsmath,amssymb,floatfix]{revtex4-1}

\usepackage{graphicx}
\usepackage{dcolumn}
\usepackage{bm}
\usepackage{multirow}

\usepackage{subfigure}

\graphicspath{{images/}{plots/}}

\newcommand{\avg}[1]{\ensuremath{\left< #1 \right>}}

\newcommand{\abs}[1]{\ensuremath{\left\vert#1\right\vert}}
\newcommand{\brac}[1]{\ensuremath{\left(#1\right)}}
\newcommand{\duck}[1]{\ensuremath{\left<#1\right>}}

\newcommand{\deff}{\ensuremath{d_{\mathrm{e}}}}

\renewcommand{\vec}[1]{\bm{#1}}

\DeclareMathOperator{\ee}{e}
\DeclareMathOperator{\conv}{conv}

\DeclareMathOperator{\Var}{Var}

\begin{document}

    \title{Convex Hulls of Random Walks in Higher Dimensions: A Large Deviation Study}
    \author{Hendrik Schawe}
    \email{hendrik.schawe@uni-oldenburg.de}
    \affiliation{Institut f\"ur Physik, Universit\"at Oldenburg, 26111 Oldenburg, Germany}
    \affiliation{LPTMS, CNRS, Univ.~Paris-Sud, Universit\'e Paris-Saclay, 91405 Orsay, France}
    \author{Alexander K. Hartmann}
    \email{a.hartmann@uni-oldenburg.de}
    \affiliation{Institut f\"ur Physik, Universit\"at Oldenburg, 26111 Oldenburg, Germany}
    \affiliation{LPTMS, CNRS, Univ.~Paris-Sud, Universit\'e Paris-Saclay, 91405 Orsay, France}
    \author{Satya N. Majumdar}
    \email{satya.majumdar@u-psud.fr}
    \affiliation{LPTMS, CNRS, Univ.~Paris-Sud, Universit\'e Paris-Saclay, 91405 Orsay, France}
    \date{\today}

    \begin{abstract}
        The distribution of the hypervolume $V$ and surface $\partial V$ of convex hulls of
        (multiple) random walks in higher dimensions are determined numerically, especially
        containing probabilities far smaller than $P = 10^{-1000}$ to estimate
        large deviation properties. For arbitrary dimensions and large walk
        lengths $T$, we suggest a scaling behavior of the distribution with the
        length of the walk $T$ similar to the two-dimensional case,
        and behavior of the distributions in the
        tails. We underpin both with numerical data in $d=3$ and $d=4$ dimensions.
        Further, we confirm the analytically known means of those
        distributions and calculate their variances for large $T$.
    \end{abstract}

    \pacs{02.50.-r, 75.40.Mg, 89.75.Da}

    \maketitle

    \section{Introduction}
        The random walk (RW) is first mentioned~\cite{hughes1996random}
        with this name in 1905 by Pearson~\cite{pearson1905problem} as
        a model, where at discrete times, steps of a fixed length are
        taken by a single walker in a random direction, e.g., with a
        random angle on a plane in two dimensions.
        This was later generalized to random flights in three dimensions~\cite{rayleigh1919xxxi}
        and RWs on a lattice in $d$ dimensions~\cite{polya1921}.
        A few decades later even more generalized models appeared,
        e.g., introducing correlation~\cite{Patlak1953random,Kareiva1983analyzing,bovet1988spatial}
        or interaction with its past trajectory~\cite{Madras2013,Lawler1980Self,weinrib1985kinetic},
        its environment~\cite{Smoluchowski1916brownsche,Alt1980biased,vanHaastert2007biased,Weesakul1961random,kac1947random}
        or other walkers~\cite{Fisher1984walks,Schehr2008Exact}.
        Despite the plethora of models developed for different applications,
        still simple isotropic RWs are used
        as an easy model for Brownian motion and diffusion processes~\cite{Smoluchowski1916brownsche,kac1947random,witten1983diffusion},
        motion of bacteria~\cite{Schaefer1973dynamics,Codling2008random},
        financial economics~\cite{Fama1965Random},
        detecting community structures in (social) networks~\cite{Rosvall2008Maps,Gupta2013WTF},
        epidemics~\cite{Dumonteil2013spatial},
        polymers in solution~\cite{Kuhn1934gestalt,helfand1975theory,Haber2000shape}
        and home ranges of animals~\cite{Bartumeus2005animal,Borger2008general}.

        The most important quantity that characterizes RWs is the
        end-to-end distance
        and how it scales with the number of steps, giving rise to an exponent $\nu$, i.e., the
        inverse fractal dimension. To describe the nature of different RW models
        more thoroughly, other quantities can be used. Here, we are interested in analyzing
        the ``volume'' and the ``surface'' of the RW, which can be conveniently
        defined by the corresponding quantities of the convex hulls of each given RW.
        These quantities are used, usually in two dimensions, to describe
        home ranges of animals~\cite{worton1995convex, Giuggioli2011animal}.
        But also, very recently, to detect different
        phases in intermittent stochastic trajectories, like the run and tumble
        phases in the movement of bacteria~\cite{grebenkov2017unraveling}.
        The convex hull of a RW
        is the smallest convex polytope containing the whole trace of the
        RW, i.e., it is a non-local characteristic that depends on the
        full history of the walker, namely all visited points.

        The most natural statistical observables associated to the convex
        hull of a random trajectory are its (hyper-) volume
        and its (hyper-) surface. The full statistics of these two random
        variables are nontrivial to compute even for a
        single Brownian motion in two or higher dimensions.
        Even less is known on the statistics of these two random variables
        for a discrete-time random walk with a symmetric and continuous jump
        distributions. In fact, most publications concentrate on the area and
        perimeter of convex hulls for two-dimensional RWs.
        The mean perimeter and the mean area of a single random walk
        in a plane, as a function of the number of steps (in the limit of large
        number of steps with finite
        variance of step lengths where it converges to a Brownian motion),
        are known exactly since more than 20
        years~\cite{Letac1980Expected,Letac1993explicit}.
        These results for the convex hull of a single Brownian motion in a plane
        have recently been generalized in several directions
        in a number of studies. These include
        the exact results for the mean perimeter and mean area of the convex hull
        for multiple independent Brownian motions and Brownian bridges in a
        plane~\cite{Majumdar2009Convex,Majumdar2010Random},
        for the mean perimeter of the convex hull of a single Brownian motion confined to a
        half plane~\cite{Chupeau2015Convex}, and for
        the mean volume and surface of the convex polytopes in arbitrary dimensions
        $d$ for a single Brownian motion and
        Brownian bridge~\cite{Eldan2014Volumetric,kabluchko2016intrinsic,vysotsky2015convex}.
        Much less is known for discrete-time random walks
        with arbitrary jump length distributions.
        Very recently the mean perimeter of the convex hull for
        planar walks for finite (but large) walk lengths and arbitrary jump
        distributions were computed explicitly~\cite{grebenkov2017mean}.
        For the special case of Gaussian jump lengths, an exact
        combinatorial formula for the
        mean volume of the convex hull in $d$-dimensions was recently
        derived~\cite{kabluchko2016intrinsic}. In $d=2$, the asymptotic
        (for large number of steps) behavior of the mean area for Gaussian
        jump lengths was derived independently in Ref.~\cite{grebenkov2017mean}.
        Also the convex hulls of other stochastic processes like L{\'e}vy
        flights~\cite{kampf2012convex,lukovic2013area},
        random acceleration processes~\cite{reymbaut2011convex} or
        branching Brownian motion with absorption~\cite{Dumonteil2013spatial}
        were under scrutiny recently.

        Analytical calculations of the variance or higher moments
        turned out to be much more difficult~\cite{snyder1993convex,Goldman1996}.
        In absence of any analytical result for
        the full distribution of the volume and surface of the convex hull of a
        random walk, a sophisticated
        large-deviation algorithm was recently used to
        compute numerically the full distribution of the perimeter and the
        area of the convex hull of a single~\cite{Claussen2015Convex} and
        multiple~\cite{Dewenter2016Convex} random walks in two dimensions. Amazingly, this
        numerical technique was able to resolve the probability
        distribution down to probabilities
        as small as, e.g., $10^{-300}$~\cite{Claussen2015Convex,Dewenter2016Convex}.
        In this work, we will use simulations to obtain the distribution
        of the volume $V$ and surface $\partial V$ of the convex hull of a
        single random walk with Gaussian jump length distribution in dimensions
        $d \in \{3,4\}$ over a large range of its support.
        In particular, this range is large enough to include large deviations,
        here down to probability densities far smaller than $P(V) = 10^{-1000}$.
        While previous work~\cite{Claussen2015Convex,Dewenter2016Convex}
        suggested that the area and perimeter distribution obeys the large deviation
        principle in $d=2$, which was later proven for the perimeter~\cite{akopyan2016large},
        our results suggest that the same holds for higher dimensions.
        Regarding the scaling behavior of the mean and of the variance, we also study higher dimensions up to $d=6$.
        Also we generalize scaling arguments to higher dimensions which were
        previously used to estimate the properties of these distributions for
        $d=2$~\cite{Claussen2015Convex}.

        The remainder of the paper is organized as follows. We will first introduce the RW model, give an
        overview for the calculation of convex hulls in higher dimensions
        and describe the sampling technique used to reach the regions of
        sufficiently small probabilities in Sec.~\ref{sec:mm}.
        The presentation of our results is split into two parts.
        Sec.~\ref{sec:means} compares our numerically obtained means
        with the analytically derived values from Refs.~\cite{Eldan2014Volumetric,kabluchko2016intrinsic}
        to check that our results are consistent with the literature.
        Also values for the variances for single and multiple
        RWs are presented. The behavior of the distributions, especially
        in their tails, is presented in Sec.~\ref{sec:distributions}.
        Sec.~\ref{sec:conclusion} concludes and gives a small outlook to
        still open questions.

    \section{Models and Methods}\label{sec:mm}
    \subsection{Random Walks}
\begin{figure}[bhtp]
            \centering
            \subfigure[\label{fig:randomWalk:2d}~$d=2, T=2048$]{
                \includegraphics[width=0.22\textwidth]{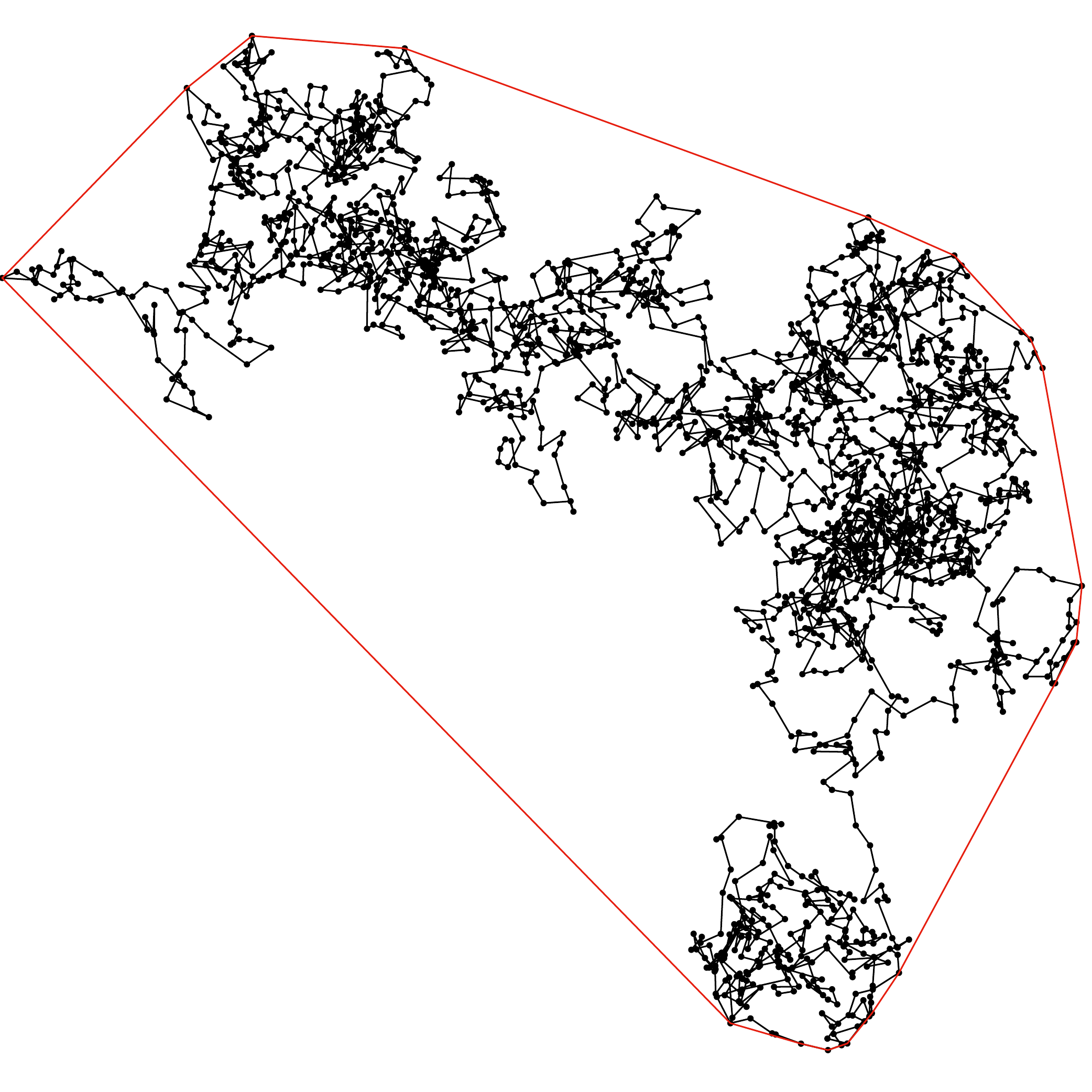}
            }
            \subfigure[\label{fig:randomWalk:3d}~$d=3, T=2048$]{
                \includegraphics[width=0.17\textwidth]{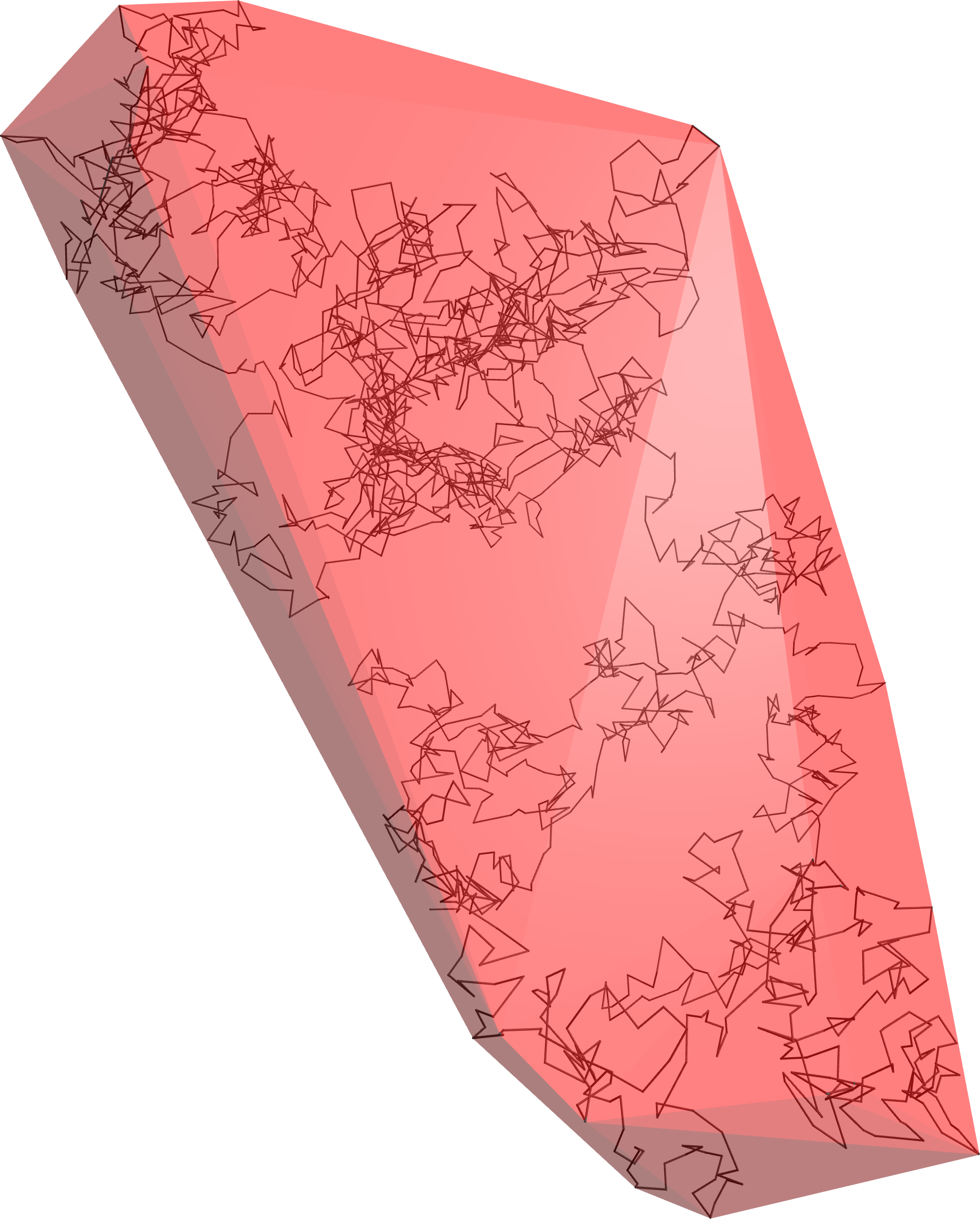}
            }
            \caption{\label{fig:randomWalk}
                (color online)
                Examples for Gaussian random walks in $d = 2$ and $d=3$.
                Their convex hull is visualized in red.
            }
        \end{figure}

        A \emph{random walk}~\cite{pearson1905problem,polya1921}
        in $d$ dimensions consists of $T$ step vectors $\vec{\delta}_{i}$
        such that its position at time $\tau$ is given as
        \begin{align*}
            \vec{x}(\tau) = \vec{x}_{0} + \sum_{i=1}^{\tau} \vec{\delta}_{i},
        \end{align*}
        where $\vec{x}_{0}$ is the starting position and chosen in the following
        always as the origin of the coordinate system.
        Thus, a realization of a walk can be characterized
        as a tuple of the displacements $(\vec{\delta}_{1}, \ldots, \vec{\delta}_{T})$.
        We will denote the set of visited points as
        $\mathcal{P} = \{\vec{x}(0), \ldots, \vec{x}(T)\}$.
        We draw the steps $\vec{\delta}_{i}$ from an uncorrelated
        multivariate Gaussian distribution with zero mean and unit width $G(0,1)$, i.e.,
        $d$ independent random numbers per step.
        Two examples for dimensions $d=2$ and $d=3$
        are visualized in Fig.~\ref{fig:randomWalk}.
        While walks on a lattice show finite-size effects of the lattice
        structure \cite{Claussen2015Convex},
        especially in the region of low probabilities, the Gaussian
        displacements lead to smooth distributions.
        Note that in the limit $T\to\infty$ Gaussian and lattice RWs do not behave differently.
        Both converge to the
        continuous-time Brownian motion~\cite{Majumdar2010Random}.

        The RW is very well investigated \cite{hughes1996random},
        especially it is known that the end-to-end distance $r$, and in fact every
        one-dimensional observable, scales as $r \propto T^{\nu}$ with $\nu=1/2$.
        This exponent $\nu$ is the same in any dimension and characteristic for
        diffusion processes.

    \subsection{Convex Hulls}
        For a given point set $\mathcal P$ its \emph{convex hull}
        $\mathcal C = \conv(\mathcal{P})$ is the
        smallest convex polytope enclosing all points $P_{i} \in \mathcal{P}$,
        i.e., all points $P_{i}$ lie inside the polytope and all straight line
        segments $(P_{i}, P_{j})$ lie inside the polytope. In Fig.~\ref{fig:randomWalk}
        two examples for $d = 2$ and $d=3$ are shown.

        Convex hulls are a well studied problem with applications from pattern
        recognition \cite{duda2012pattern} to ecology studies~\cite{Cornwell2006Trait}.
        They are especially important
        in the context of computational geometry,
        where next to a wide range of direct applications~\cite{Preparata1985convex,jayaram2016convex}
        the construction of Voronoi diagrams and Delaunay triangulations~\cite{brown1979Voronoi}
        stand out, which in turn are useful in a wide range of disciplines~\cite{Aurenhammer1991Voronoi}.
        Note that a lower bound for the worst-case time complexity of an
        exact convex hull algorithm for $T = \abs{\mathcal P}$ points is
        $\Omega\brac{T^{\left \lfloor d/2 \right \rfloor}}$~\cite{klee1980complexity, seidel1981convex, clarkson1989applications},
        which is the order of possible facets, i.e., exponential in the dimension.
        Although, there are approximate algorithms~\cite{Xu1998approximate,Sartipizadeh2016computing}
        which probably would make the examination of higher dimensional
        convex hulls feasible, we are only examining the convex hulls up to
        $d=6$ using exact algorithms.

\begin{figure}[hbtp]
            \centering
            \subfigure[\label{fig:simplexVolume}]{
                \includegraphics[width=0.14\textwidth]{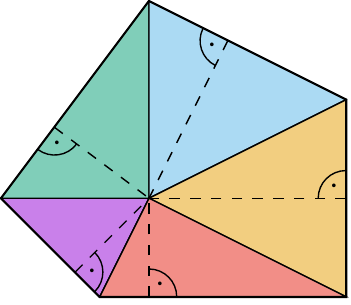}
            }
            \subfigure[\label{fig:quickhull_2}]{
                \includegraphics[width=0.14\textwidth]{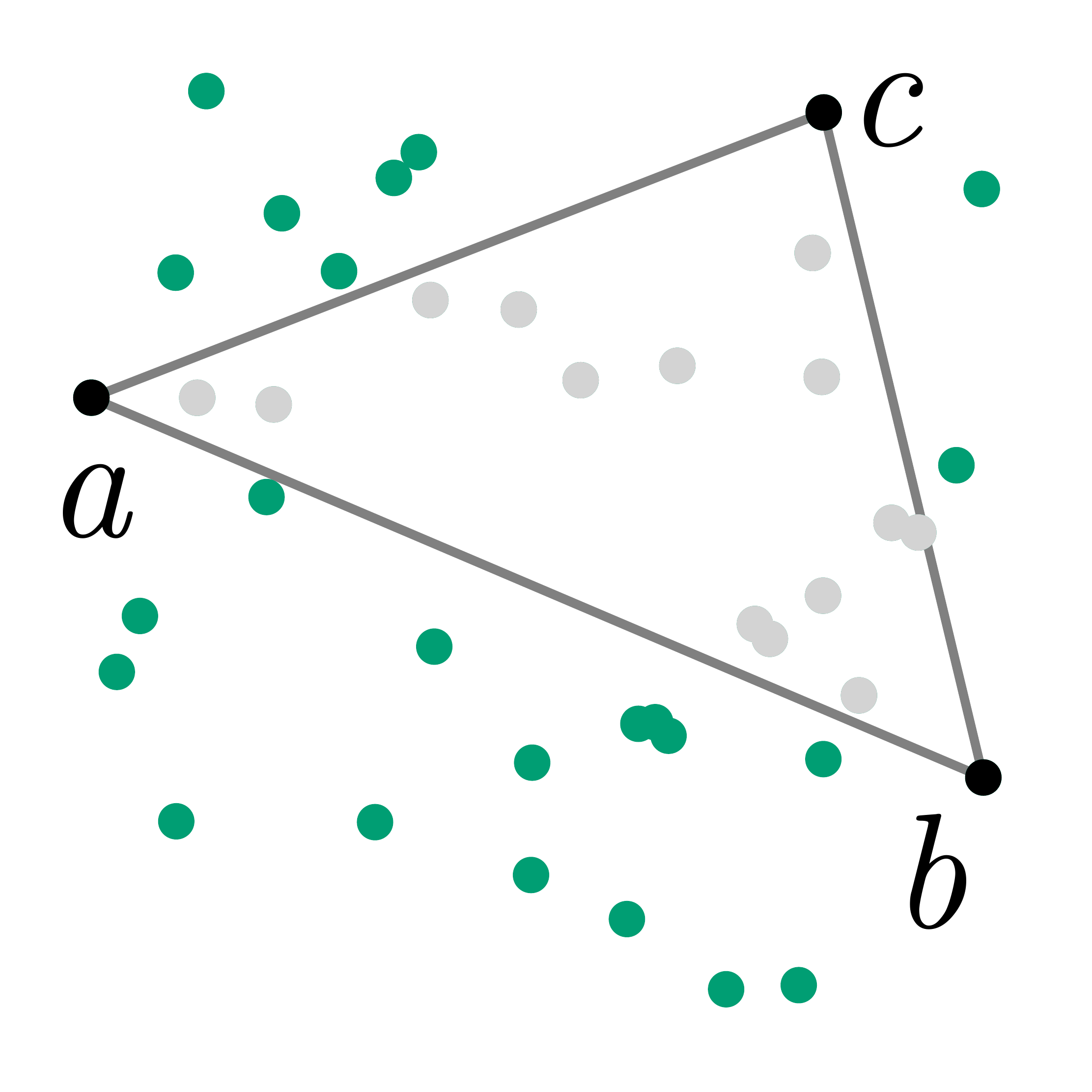}
            }
            \subfigure[\label{fig:quickhull_3}]{
                \includegraphics[width=0.14\textwidth]{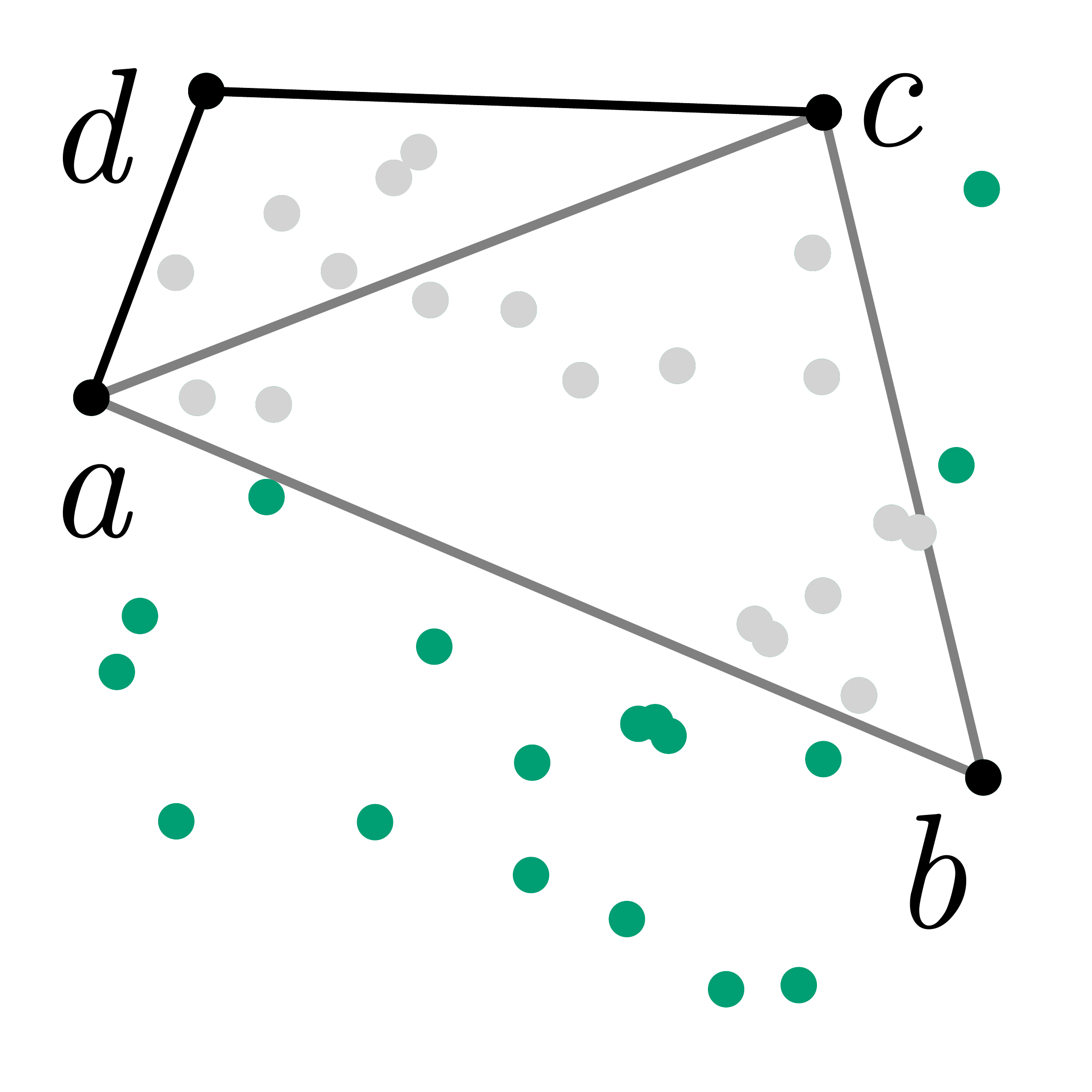}
            }
            \caption{\label{fig:quickhull}
                (color online)
                Visualization of \subref{fig:simplexVolume} the idea to
                calculate the volume of a convex polygon given its facets
                and an interior point, perpendicular distances are shown with
                dashed lines. In
                \subref{fig:quickhull_2} and \subref{fig:quickhull_3} examples of two
                consecutive recursive steps of the quickhull
                algorithm are shown. The point $d$ is left of and farthest away
                from $(a,c)$. Parts of the convex
                hull are black, discarded points are light gray.
            }
        \end{figure}

        We measure the \emph{(hyper-) volume} $V$, e.g., in $d=3$ the volume, and the
        \emph{(hyper-) surface}  $\partial V$, e.g., in $d=3$ the surface area.
        Determining surface and volume of a high-dimensional convex
        polytope is trivial given its facets $f_i$, which are ($d-1$)-dimensional
        simplexes.
        Choosing an arbitrary fixed point $p$ inside the convex polygon, one can create a
        $d$-dimensional simplex from each facet $f_i$, such that their union
        fills the entire convex hull (cf.~Fig.~\ref{fig:simplexVolume} for a
        $d=2$ example). Therefore the volume can be obtained by calculating
        \begin{align*}
            V = \sum_i \mathrm{dist}(f_i, p) a_i / d,
        \end{align*}
        where $\mathrm{dist}(f, p)$ is the perpendicular distance from the facet
        $f_i$ to the point $p$ and $a_i$ is the surface of the facet.
        The surface of a $(d-1)$-dimensional facet is its $(d-2)$-dimensional volume,
        which can be calculated with the same method recursively, until the
        trivial case of one dimensional facets, i.e., lines.
        Determining the surface uses the same recursion, by calculating
        $\partial V = \sum_i a_i.$

        To foster intuition, this method is pictured for $d=2$ in
        Fig.~\ref{fig:simplexVolume}. Here, the facets are lines and the volume
        of the simplex is the area of the triangle. The perpendicular distances
        are visualized as dashed lines.

        In the scope of this
        study, we use the \emph{quickhull} algorithm~\cite{Eddy1977Convex,Bykat1978Convex,Mucke2009Quickhull},
        and its excellent implementation in the \emph{Qhull} library~\cite{Barber1996thequickhull}.
        Quickhull is a divide-and-conquer algorithm applicable in arbitrary
        dimensions.
        For clarity, the algorithm will be explained for $d=2$, since it
        makes the central idea clear. The technical details and the
        generalization to higher dimensions are well explained in Ref.~\cite{Barber1996thequickhull}.

        Start with two points $a, b$ on the convex hull, e.g., the points
        with minimum and maximum $x$-coordinate. Determine the point $c$
        left (when ``looking'' $a\to b$) of and farthest away from the edge $(a, b)$
        and discard all points inside the polygon $(a, b, c)$. Repeat this
        step recursively with the edges $(a, c)$ and $(c, b)$
        until there are no points on the left side of the current edge.
        All edges created in this way on the bottom level of the recursion are
        part of the convex hull. Two steps of this
        recursion are pictured in Fig.~\ref{fig:quickhull}\subref{fig:quickhull_2}\subref{fig:quickhull_3}.
        The same process is repeated recursively
        with the point $c'$ left  of and farthest away from the inverse edge $(b, a)$.

    \subsection{Sampling}
        We performed Markov chain Monte Carlo simulations to examine the distributions of
        the volume $V$ and the surface $\partial V$ of the convex hull
        of RWs in dimensions $d \in \{3, 4\}$. To collect \emph{large-deviation}
        statistics, i.e., obtain not only the peak, but also
        the tails of the distribution, we use both the classic \emph{Wang Landau} (WL)
        sampling~\cite{Wang2001Efficient, Wang2001Determining}  and a
        modified Wang Landau sampling~\cite{Schulz2003Avoiding,Belardinelli2007Fast,Belardinelli2007theoretical}
        with a subsequent entropic sampling~\cite{Lee1993Entropic,Dickman2011Complete}
        run. In contrast to similar studies~\cite{Claussen2015Convex,Dewenter2016Convex}
        no temperature-based sampling scheme was used, since the difficulties
        to find suitable temperatures and regarding equilibration mentioned
        in Ref.~\cite{Claussen2015Convex} are even worse in higher dimensions.

        Both sampling techniques generate Markov chains of \emph{configurations}, where here
        configurations are realizations, each given by the tuple of RW displacements
        $(\vec{\delta}_{1}, .., \vec{\delta}_{T})$.
        One only needs  a function
        yielding an ``energy'' of a configuration
        and a way to change a
        configuration to a similar configuration. As energy we simply use
        the observable of interest $S$, i.e., either the volume $V$ or
        the surface $\partial V$. To change a configuration,
        we replace a randomly chosen step $\vec{\delta}_{i}$ of the RW  with a new
        randomly drawn step. Because all points $\vec{x}(\tau)$ for $\tau \ge i$ change,
        this is a global change of the walk. Though, this does not lead to a severe computational
        overhead, because after the update the convex hull has to be calculated
        again from scratch in any case.

        For both WL versions at first a lower and upper bound of the observable
        $S$ needs to be defined and the range in between is subdivided in
        overlapping \emph{windows}, depending on system size $T$. For the present work
        it was sufficient to sample each window independently in parallel. Therefore,
        it was not necessary to apply a replica-exchange enhancement~\cite{landau2013generic}.

        In the beginning, we start with an arbitrary configuration $c_i$ of the walk.
        Afterwards we repeatedly propose random changes each leading to a new
        configuration $c_{i+1}$ and accept each with the Metropolis acceptance
        probability
        \begin{align}
            \label{eq:wlacc}
            p_\mathrm{acc}(S(c_i) \to S(c_{i+1})) = \min\brac{\frac{g(S(c_{i}))}{g(S(c_{i+1}))}, 1},
        \end{align}
        where $g$ is an estimate for the density of states -- basically the
        wanted distribution. If $g$ equals the true density of states this will result in
        every $S$ being visited with the same probability, i.e., a flat
        histogram of $S$. Since we do not know the true density of states in advance, WL iteratively improves the
        estimate $g$. Therefore, every
        time a value of $S$ is visited, $g(S)$ is increased.
        The original article suggests to multiply $g(S)$ with a fixed factor $f$
        to perform the increase, i.e., $g(S) \mapsto g(S) f$,
        and after an auxiliary histogram fulfills some flatness criterion
        reduce this factor $f \mapsto \sqrt{f}$. This is repeated until $f$ falls below
        some beforehand defined threshold $f_\mathrm{final}$.
        Since the acceptance ratio changes during the simulation, detailed
        balance does not hold, such that systematic errors are introduced.
        To mitigate this, a better schedule to modify $g$ is introduced in~\cite{Belardinelli2007Fast},
        which reduces the systematic errors. Basically, the flatness criterion is
        removed and the factor by which to increase $g(S)$ when visiting $S$ is
        a function of the \emph{Monte Carlo time} $t$ of the simulation,
        i.e., $\ln(g(S)) \mapsto \ln(g(S)) + t^{-1}$. The sampling terminates
        as soon as as $t^{-1} \le f_\mathrm{final}$.
        This has the added benefit that the simulation time does not depend on
        some flatness criterion, which is hard to predict, but is at most
        $1/f_\mathrm{final}$ Monte Carlo sweeps.

        To remove the systematic error completely, one can use entropic
        sampling~\cite{Lee1993Entropic,Dickman2011Complete},
        i.e., fix the so far obtained estimate $g$ and sample the system using the
        same acceptance as before from Eq.~\eqref{eq:wlacc}.
        This obeys detailed balance. Finally, one
        creates a
        histogram $H$ of the visited $S$ to arrive at a corrected
        $\widetilde{g}(S) = g(S) H(S)/\avg{H}$~\cite{Dickman2011Complete}, where
        $\avg{H}$ is the average number of counts of the histogram.

        During this simulation, the value $S$ of the configuration may not leave its window, thus
        changes to configurations outside of the window are rejected.
        This also means that the first configuration must be within the
        window and is therefore obtained via a greedy heuristic.
        The final distribution is obtained as follows: For mutually overlapping windows, the corresponding
        densities are multiplied by factors such that in the overlapping regions the densities agree as much
        as possible. Finally the density obtained in this way is normalized yielding the whole distribution.
        To estimate the errors of the distribution, this simulation is done
        a couple of times and the standard error of the single bins is used as
        an error estimate.

        For the results, which  we will present in the following section, we used
        data from both sampling techniques and in some cases merged
        them. Comparisons of both techniques showed that the errors introduced
        by WL have no considerable influence on our results (not shown).

        For the determination of mean and variances of convex hull volume and surface the contribution of
        the tails are negligible, thus we used simple sampling, which
        enables the simulation of longer walks, i.e., larger values of $T$, in
        a larger range of dimensions $d=2, \ldots, 6$,

    \section{Results}
    \subsection{Mean and Variance}\label{sec:means}
        At first, we will verify our simulations by comparing with some analytically known
        results~\cite{Letac1980Expected,Eldan2014Volumetric}
        for the mean volume $V$ and surface $\partial V$ scaled
        appropriately as $\mu_V = \duck{V} / T^{d\nu}$ and $\mu_{\partial V} = \duck{\partial V} / T^{(d-1)\nu}$.
        The scaling comes from the $r\propto T^{\nu}$ scaling of the RW end-to-end distance in combination
        with the typical scaling $V\propto r^d$ and $\partial V \propto r^{d-1}$.
        For large $T$ it is known that
        \begin{align}
            \mu_{V}^{\infty} &= \brac{\frac{\pi}{2}}^{d/2} \Gamma\brac{\frac{d}{2}+1}^{-2}, \\
            \mu_{\partial V}^{\infty} &= \frac{2{(2\pi)}^{(d-1)/2}}{\Gamma(d)}.
        \end{align}
        This simulation uses simple sampling to sample $10^6$ (fewer for
        $d = 6$ resulting in larger uncertainties) sufficiently long
        walks of up to $T=262144$.

        \begin{figure}[bhtp]
            \centering
            \includegraphics[scale=1]{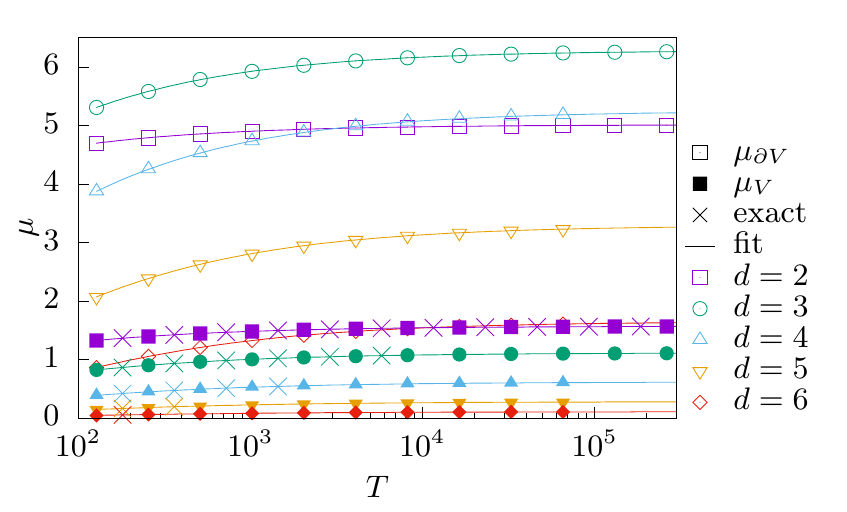}
            \caption{\label{fig:means}
                (color online)
                Scaled mean of the surface $\mu_{\partial V} = \duck{\partial V} / T^{(d-1)\nu}$
                (open symbols) and volume $\mu_{V} = \duck{V} / T^{d\nu}$ (solid
                symbols) for different dimensions (different shapes) and walk
                lengths $T$ obtained by $10^{6}$ samples each. Lines are
                fits (cf.~Eq.~\eqref{eq:mean}) to extrapolate for $T \to \infty$.
                Crosses are exact values (cf.~Eq.~\eqref{eq:exactMeansT}) and
                show very good agreement with the extrapolation.
                The asymptotic values are shown in Tab.~\ref{tab:measuredMu}.
                Fit ranges: $d \le 4$: $T\ge128$, $d \ge 5$: $T\ge256$
                for the surface and
                $d \le 4$: $T \ge 128$, $d \ge 5$: $T\ge512$ for the volume
                (same ranges for the variances).
                The goodness of fit $\chi^{2}_{\mathrm{red}}$ is between $0.3$
                and $1.2$ for all fits. Errorbars are smaller than the line
                of the fit.
            }
        \end{figure}

        There is an exact result for the mean Volume of the convex hull
        for finite $T$~\cite{kabluchko2016intrinsic}:
        \begin{align}
            \label{eq:exactMeansT}
            \duck{V} = \frac{2^{-d/2}}{\Gamma(d/2+1)} \sum_{n_1,\ldots, n_d}\frac{1}{\sqrt{n_1 \ldots n_d}} I(n_1,\ldots ,n_d),
        \end{align}
        where $1 \le n_i \le T$ are integers and
        \begin{align*}
            I(n_1,\ldots ,n_d) =
            \begin{cases}
                1 \quad \text{ if } n_1 + \ldots + n_d \le T\\
                0 \quad \text{ else }.
            \end{cases}
        \end{align*}
        E.g.~for $d=2$ and $d=3$ this results in
        \begin{align}
            \duck{V_2} &= \frac{1}{2} \sum_{i=1}^T\sum_{j=1}^{T-i}\frac{1}{\sqrt{ij}}\\
            \duck{V_3} &= \frac{2^{3/2}\cdot 4}{3\sqrt{\pi}} \sum_{i=1}^T\sum_{j=1}^{T-i}\sum_{k=1}^{T-i-j}\frac{1}{\sqrt{ijk}}
        \end{align}
        respectively.
        The number of elements in the sums grows with $\mathcal{O}(T^d)$
        in the number of steps $T$ and the dimension $d$, such that
        a numerical evaluation is only feasible for rather small $T$ and $d$.
        We calculated some exact values to ensure the quality of our
        simulations and the extrapolation. These are marked with crosses in
        Fig.~\ref{fig:means}.

        To estimate the $T \to \infty$ asymptotic
        value $\mu_{V}^\infty$, it is necessary to extrapolate measurements
        for different lengths $T$.
        We fit the expansion
        \begin{align}
            \label{eq:mean}
            \duck{V} / T^{d\nu} = \mu_V + C_1 T^{-1/2} + C_2 T^{-1}
        \end{align}
        also used in Ref.~\cite{grebenkov2017mean} to our measurements.
        This produces very good fits, shown in Fig.~\ref{fig:means}, and
        values in very good agreement with the expectations.
        We use the same function for the surface and the variances.
        Though small values of $T$ need to be excluded from the fits,
        especially for high dimensions. The precise fit ranges are listed in
        the caption of Fig.~\ref{fig:means}.

        \begin{table}[htb]
            \begin{ruledtabular}
                \begin{tabular}{rlllll}
                    \multicolumn{1}{c}{$d$} & \multicolumn{1}{c}{$\mu_V^\infty$} & \multicolumn{1}{c}{$\mu_{\partial V}^\infty$} & \multicolumn{1}{c}{${\sigma_V^\infty}^2$} & \multicolumn{1}{c}{${\sigma_{\partial V}^\infty}^2$} & \multicolumn{1}{c}{$\frac{\sigma_{V}^\infty}{\mu_{V}^\infty}$}\\[0.05cm]
                    \hline
                    \noalign{\vskip 0.1cm}
                    2 & 1.5708    & 5.0132    &            &          &           \\
                    3 & 1.1140    & 6.2832    &            &          &           \\
                    4 & 0.6168    & 5.2499    &            &          &           \\
                    5 & 0.2800    & 3.2899    &            &          &           \\
                    6 & 0.1077    & 1.6493    &            &          &           \\
                    \hline
                    \noalign{\vskip 0.1cm}
                    2 & 1.5705(3) & 5.0127(5) & 0.3078(3)  & 1.077(1) & 0.3532(2)\\
                    3 & 1.1139(2) & 6.2832(9) & 0.1778(2)  & 3.093(3) & 0.3785(2)\\
                    4 & 0.6164(1) & 5.2473(10)& 0.05882(7) & 2.808(3) & 0.3932(2)\\
                    5 & 0.2801(1) & 3.2909(9) & 0.01274(2) & 1.279(2) & 0.4032(3)\\
                    6 & 0.1077(1) & 1.6492(6) & 0.00193(1) & 0.351(1) & 0.4080(5)\\
                \end{tabular}
            \end{ruledtabular}
            \caption{\label{tab:measuredMu}
                Analytically expected (top, rounded to four decimal places) and
                from measurements extrapolated (bottom) asymptotic mean and
                variance of volume, respectively surface.
                Analytical values for the variances are unknown
                (except for Brownian bridges~\cite{Goldman1996spectrum}).
                Though for the perimeter ($d=2$) rigorous bounds~\cite{wade2015convex} are known
                ${\sigma_{\partial V}^\infty}^2 \in [2.65 \cdot 10^{-3}, 9.87]$
                Error estimates for the last column are obtained by Gaussian error
                propagation.
            }
        \end{table}

        The obtained asymptotic values are listed in Tab.~\ref{tab:measuredMu}.
        Mind, that the error estimates are purely statistical and do not
        take into
        account higher order terms than those present in Eq.~\eqref{eq:mean}.
        To make matters worse, not the same large system sizes could be reached for
        higher dimensions due to the exponentially increasing time complexity~\cite{klee1980complexity}.

        Also, we looked at the average volume $\mu_V = \left< V \right>/T^{d\nu}$
        and surface $\mu_{\partial V} = \left< \partial V \right>/T^{(d-1)\nu}$
        of the convex hulls of multiple RWers with $n \in \{2,3,10, 100\}$
        independent RWs in $d=3$ dimensions, which are tabulated in
        Tab.~\ref{tab:measuredMuMulti}.
        We determined the listed values in the same way as before with a fit to
        Eq.~\eqref{eq:mean} (no figure shown) within the same ranges as single
        walks.

        \begin{table}[htb]
            \begin{ruledtabular}
            \begin{tabular}{rllll}
                \multicolumn{1}{c}{$n$} & \multicolumn{1}{c}{$\mu_V^\infty$} & \multicolumn{1}{c}{$\mu_{\partial V}^\infty$} & \multicolumn{1}{c}{${\sigma_V^\infty}^2$} & \multicolumn{1}{c}{${\sigma_{\partial V}^\infty}^2$}\\[0.05cm]
                \hline
                \noalign{\vskip 0.1cm}
                  2 & \phantom{1}3.151    & 12.566                &           &          \\[0.06cm]
                \hline
                \noalign{\vskip 0.1cm}
                  2 & \phantom{1}3.153(1) & 12.572(2)  & \phantom{1}1.427(1)  & 12.40(1) \\
                  3 & \phantom{1}5.332(1) & 17.644(2)  & \phantom{1}3.796(4)  & 21.66(2) \\
                 10 &           17.695(2) & 37.528(3) &           22.54(3)    & 48.65(4) \\
                100 &           66.233(7) & 85.563(5) &           68.65(10)   & 56.44(7) \\
            \end{tabular}
            \end{ruledtabular}
            \caption{\label{tab:measuredMuMulti}
                Analytically expected (top) and from measurements
                extrapolated (bottom) mean and variance
                of the volume, respectively surface of the convex hull
                of $n$ independent RWs in $d=3$ dimensions.
                Analytical values for the variances are unknown.
                The quality of fit $\chi^2$ for all fits is between $0.4$
                and $1.7$.
            }
        \end{table}

        Since the single steps $\vec{\delta}_i$ are independent,
        two walkers, i.e., the $n=2$ case, can be joined at the origin to one
        walk with twice the number of steps~\cite{engel2016disputation}, thus
        $\mu_{V_2}^\infty = 2^{d\nu} \mu_V^\infty$ and $\mu_{\partial V_2}^\infty = 2^{(d-1)\nu} \mu_{\partial V}^\infty$
        are the exact mean values for this case.
        The numerical data is within statistical errors compatible
        with this expectation.
        Though, for $n>2$ this is not as easy anymore. We are
        not aware of any other published expectations for $d\ge3$.

        We have performed the same analysis (no figure shown) for the variances
        $\sigma_V^2 = \Var{(V)} / T^{2d\nu}$ and
        $\sigma_{\partial V}^2 = \Var{(\partial V)} / T^{2(d-1)\nu}$
        and the same remarks apply.

        For the ratio between standard deviation and mean
        \begin{align*}
            \lim_{d\to \infty} \frac{\sigma_V^\infty}{\mu_V^\infty} = 0
        \end{align*}
        is conjectured~\cite{Eldan2014Volumetric}.
        Our data shows no downward trend for this ratio as shown in the last
        column of Tab.~\ref{tab:measuredMu}. However, to draw any
        conclusions, one should gather results for $d \gg 6$, which may be
        possible using some fast approximation scheme for convex hulls in high
        dimensions, though this is beyond the scope of this study.

        \begin{figure}[bhtp]
            \includegraphics[scale=1]{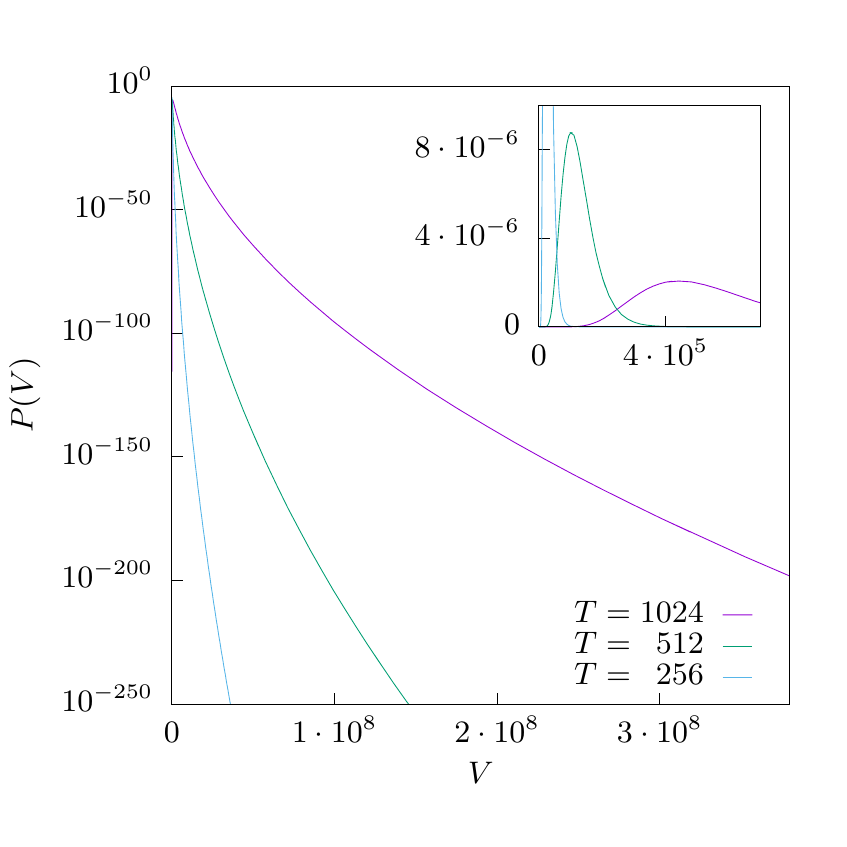}
            \caption{\label{fig:unscaled}
                (color online)
                Distribution of the volume of a $d=4$ RW for
                different system sizes $T$. The inset shows the peak region
                in linear scale.
            }
        \end{figure}

        \begin{figure*}[bhtp]
            \centering
            \subfigure[\label{fig:scaling:GW3Ddel}~$d=3, \widetilde{S}>500, b_\mathrm{r}=1.55, \chi^2_{\mathrm{red}} = 2.5$]{
                \includegraphics[scale=1]{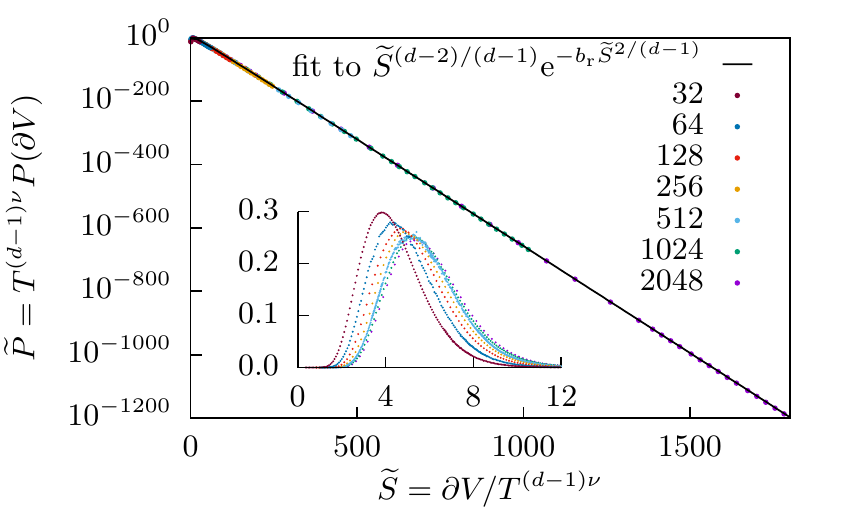}
            }
            \subfigure[\label{fig:scaling:GW4Ddel}~$d=4, \widetilde{S}>200, b_\mathrm{r}=6.33, \chi^2_{\mathrm{red}} = 1.2$]{
                \includegraphics[scale=1]{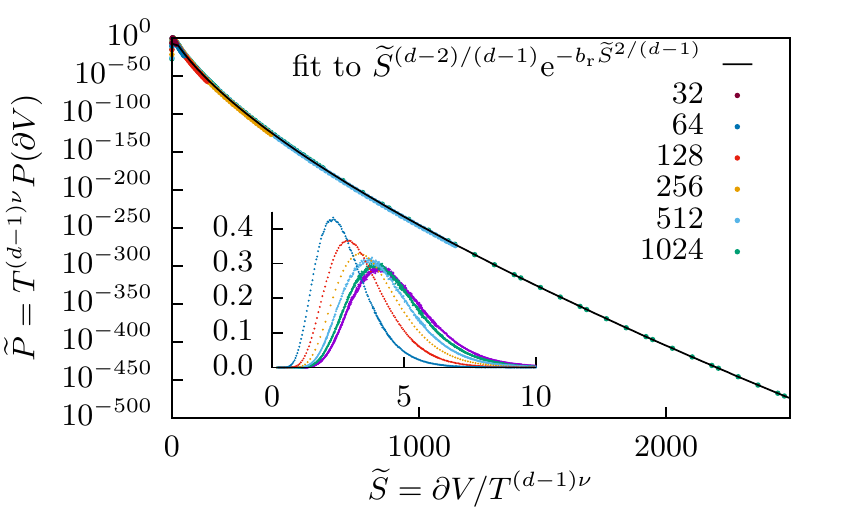}
            }

            \subfigure[\label{fig:scaling:GW3D}~$d=3, \widetilde{S}>500, b_\mathrm{r}=10.61, \chi^2_{\mathrm{red}} = 0.8$]{
                \includegraphics[scale=1]{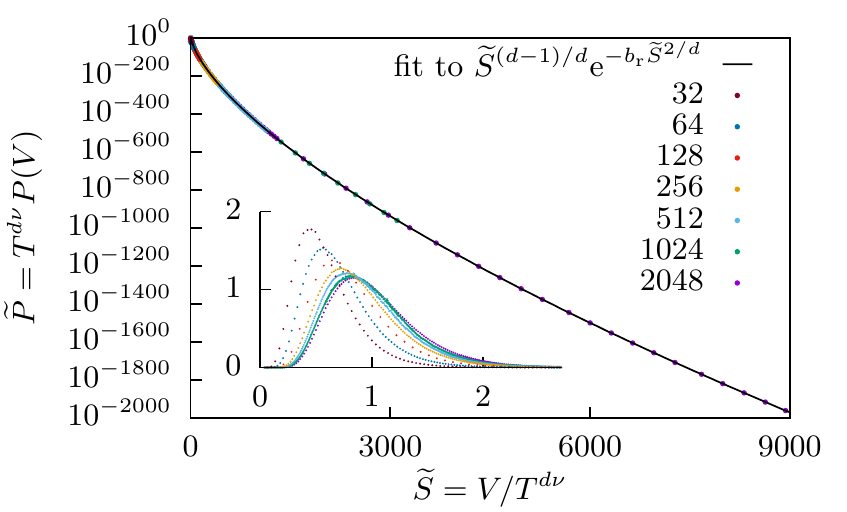}
            }
            \subfigure[\label{fig:scaling:GW4D}~$d=4, \widetilde{S}>2500, b_\mathrm{r}=26.61, \chi^2_{\mathrm{red}} = 1.1$]{
                \includegraphics[scale=1]{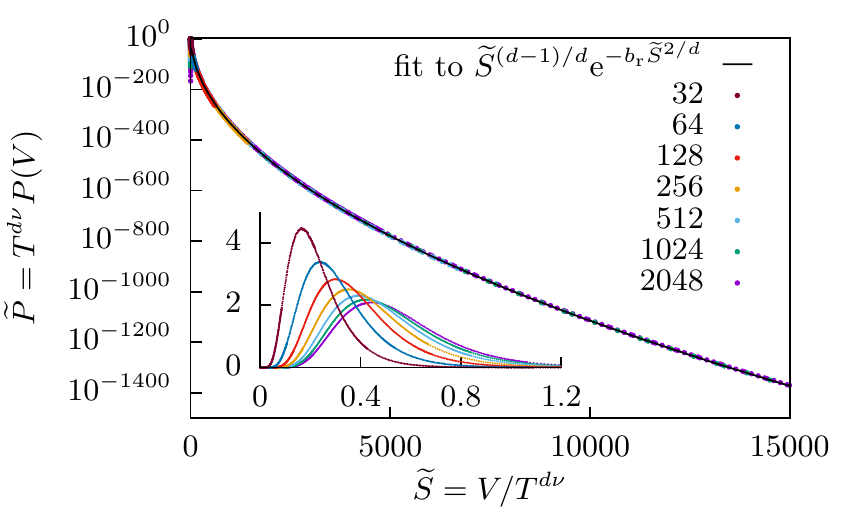}
            }
            \caption{\label{fig:scaling}
                (color online)
                Distributions of the surface (top) and volume (bottom)
                for $d\in\{3,4\}$ scaled according to Eq.~\eqref{eq:scaling}.
                Statistical errors are smaller than the symbols.
                The scaling indeed collapses the distributions on one scaling function
                $\widetilde{P}$. Fits are shown for the largest system size.
                The inset shows the peak region in linear scale. For larger
                values of $T$ the collapse works better.
                (Only a small fraction of all data points are visualized.)
            }
        \end{figure*}

    \subsection{Distributions}\label{sec:distributions}
        In addition to the first moments shown in the previous section,
        here we look at the actual distribution over a large part of
        the support. Since the Gaussian distribution, from which the steps
        are drawn, is not bounded, $V$ and $\partial V$ of
        a walk consisting of such steps are not bounded, either. Therefore,
        not the whole support, but a reasonably large part is sampled.
        Especially, it is large enough to investigate the large-deviation properties of the
        distribution. As an example, a part of the distribution
        for the volume of a convex hull of RWs in $d=4$
        dimensions is shown in Fig.~\ref{fig:unscaled}.

        As we mentioned in the previous section, $\avg{V}$ and $\avg{\partial V}$
        scale for large values of $T$ as $T^{\deff\nu}$ where $\deff$
        is the effective dimension of the observable, i.e., $\deff = d$ for the
        volume and $\deff = d-1$ for the surface. A natural question is,
        if the whole distribution does scale according to $T^{\deff\nu}$.
        Ref.~\cite{Claussen2015Convex} already shows that this is true for $d=2$.
        For higher dimension we arrive analogously at the scaling assumption
        for the distribution of the observable $S$
        \begin{align}
            \label{eq:scaling}
            P(S) = T^{-\deff\nu} \widetilde{P}(S T^{-\deff\nu}).
        \end{align}
        Fig.~\ref{fig:scaling} shows the distributions of the volume and
        surface of the convex hulls of RWs in $d \in \{ 3,4 \}$ dimensions
        scaled according to Eq.~\eqref{eq:scaling}.
        Apparently the scaling works very well in the right tail of larger
        than typical $V$. The inset shows that in the peak region there
        are major corrections to the assumed scaling for small values of $T$, but
        it also shows that those corrections rapidly get smaller for
        larger values of $T$. A power-law fit with offset to the position of
        the maxima of the distributions (no figure) with increasing walk length $T$,
        confirms convergence for large values of $T$, i.e., the peaks do collapse
        on one universal curve for $T \to \infty$.

        In fact, the scaling for the distribution of the span $s$, which
        is the distance between the leftmost and rightmost point, of a one
        dimensional Brownian motion is known~\cite{hughes1996random,Kundu2013exact}
        to be
        \begin{align*}
            P(s, T) = \brac{4DT}^{-\nu} f\brac{\frac{s}{\brac{4DT}^{\nu}}},
        \end{align*}
        with some diffusion constant $D$ and
        \begin{align*}
            f(x) = \frac{8}{\sqrt{\pi}} \sum_{m=1}^{\infty} {(-1)}^{m+1} m^{2} \ee^{-m^{2}x^{2}}
        \end{align*}
        which has the following asymptotic behavior \cite{Claussen2015Convex}:
        \begin{align*}
            f(x) &= 2\pi^{2}x^{-5}\ee^{-\pi/4x^{2}}, &&\text{for } x \to 0 \\
            f(x) &= \frac{8}{\sqrt\pi} \ee^{-x^{2}}, &&\text{for } x \to \infty
        \end{align*}
        Finally, substituting $s \propto S^{1/\deff}$ leads to a guess for
        the expected behavior of the tails with
        \begin{align}
            \label{eq:leftTail}
            \widetilde{P}(\widetilde{S}) &\propto \widetilde{S}^{(\deff-6)/\deff} \ee^{-b_\mathrm{l} \widetilde{S}^{-2/\deff}}, &&\text{for } \widetilde{S}\to 0\\
            \label{eq:rightTail}
            \widetilde{P}(\widetilde{S}) &\propto \widetilde{S}^{(\deff-1)/\deff} \ee^{-b_\mathrm{r}\widetilde{S}^{2/\deff}},                        &&\text{for } \widetilde{S}\to\infty
        \end{align}
        where a rescaled $\widetilde{S} = S T^{-\nu\deff}$ is introduced
        for clarity and with free parameters $b_\mathrm{l}$ and $b_\mathrm{r}$.
        The $\frac{\mathrm{d}s}{\mathrm{d}S} \propto S^{(\deff-1)/\deff}$ factors are
        introduced by the substitution.

        \begin{figure}[bhtp]
            \includegraphics[scale=1]{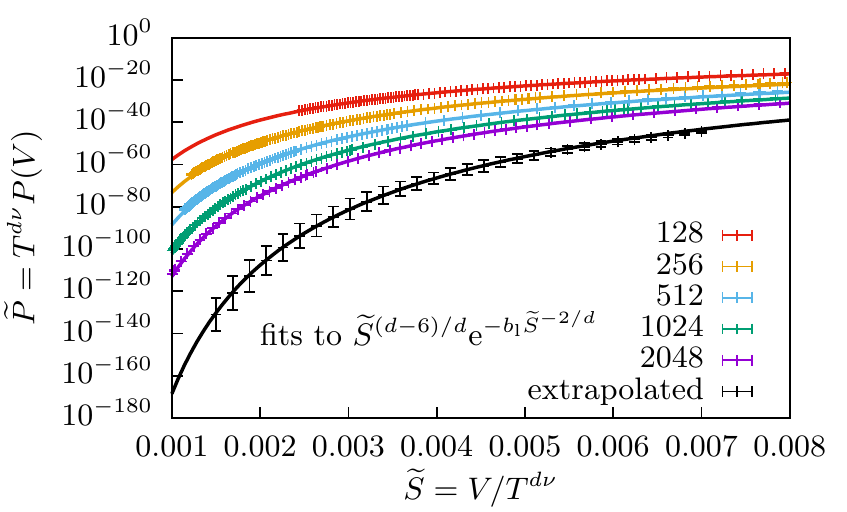}
            \caption{\label{fig:leftTail4D}
            (color online)
            Fit of the exponential Eq.~\eqref{eq:leftTail} to the left tail.
            (for clarity, only every tenths data point is plotted)
            }
        \end{figure}

        For all values of $T$, the expected distribution for the left tail Eq.~\eqref{eq:leftTail}
        fits well to the sampled data, shown for the example of the volume
        in $d=4$ in Fig.~\ref{fig:leftTail4D}.
        We extrapolated the curve point-wise to $T\to\infty$ assuming
        a power-law scaling, resulting in the limit curve in Fig.~\ref{fig:leftTail4D}.
        Similar to the main region of
        the distribution (shown in Fig.~\ref{fig:scaling}), smaller values
        of $T$ show larger deviations from the
        limit curve.  Note that also the limiting curve fits Eq.~\eqref{eq:leftTail}
        (with a suitable values for $b_\mathrm{l}$ and the prefactor).

        The same analysis for the right tails is shown in Fig.~\ref{fig:scaling},
        where Eq.~\eqref{eq:rightTail} is fitted to
        the right tail of the distributions of the volume and surface in
        $d\in \{ 3,4 \}$.
        The good $\chi^2_\mathrm{red}$ values
        suggest that this is a good estimate of the asymptotic behavior indeed.

        To determine whether a distribution $P$ satisfies the large deviation
        principle, i.e., whether it scales as
        \begin{align}
            \label{eq:rate}
            P_T \approx \ee^{-T\Phi}
        \end{align}
        for some large parameter $T$, we look if the \emph{rate function} $\Phi$
        does exist in the $T\to\infty$ limit~\cite{Touchette2009large}.
        Comparing Eq.~\eqref{eq:rate} to the behavior of the right tail (cf.~Fig.~\ref{fig:scaling}
        and Eq.~\eqref{eq:rightTail}) the rate function seems to be a power law
        with an exponent $\kappa = 2/\deff$, i.e.,
        \begin{align}
            \label{eq:rateGuess}
            \Phi(S) \propto S^{\kappa} = S^{2/\deff}
        \end{align}

\begin{figure}[bhtp]
            \includegraphics[scale=1]{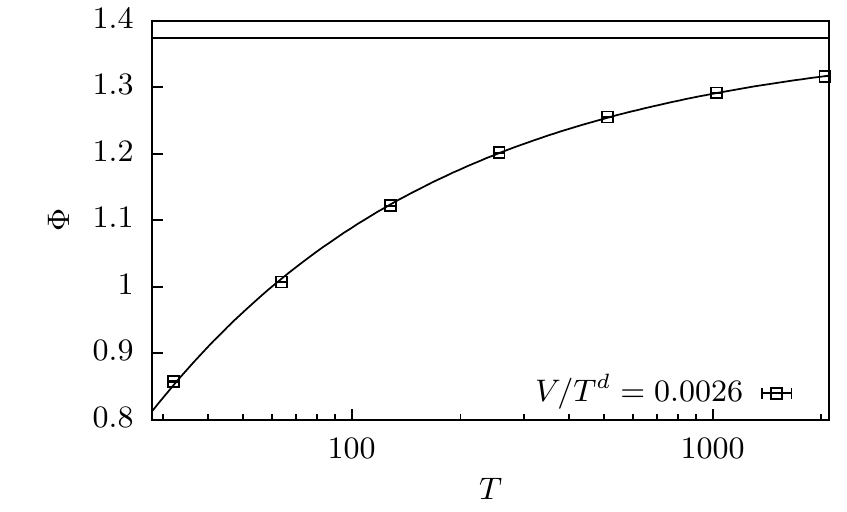}
            \caption{\label{fig:rateExtra}
                Point-wise extrapolation of the value of the rate function at a fixed value $V/T^d$
                to $T\to \infty$ with a power law,
                here for a $d=4$ dimensional volume. The power-law fit
                seems to be a reasonable approximation.
            }
        \end{figure}

        Since we have numerical results for the distribution $P$, we can
        determine an empirical rate function $\Phi$ of the volume/surface
        $S$ by extrapolation of
        \begin{align}
            \label{eq:defRateFunction}
            \Phi(S/S_{\mathrm{max}}) = -\frac{1}{T} \ln P(S/S_{\mathrm{max}})
        \end{align}
        to the large $T$ limit.
        While $\Phi$ is usually normalized to $\Phi \in [0,1]$,
        here $S$ and thus $\Phi$ is not bounded.
        To get a rate function $\Phi$ comparable to other publications,
        we assume $S_{\mathrm{max}} = T^{\deff}$ like
        Ref.~\cite{Claussen2015Convex}.
        We extrapolated the empirical rate function point-wise for $T\to\infty$
        using a power law with offset as shown in Fig.~\ref{fig:rateExtra}.
        Note that
        since for different walk lengths $T$ we used different histogram bins, we obtain the
        intermediate values between the discrete bins by cubic spline
        interpolation. The extrapolation  leads to an
        asymptotic rate function estimate. This shows that the rate function exists and
        this distributions satisfies the large-deviation principle. This holds
        for $d=3$ and $d=4$, for both volume and surface.

        \begin{figure}[bhtp]
            \includegraphics[scale=1]{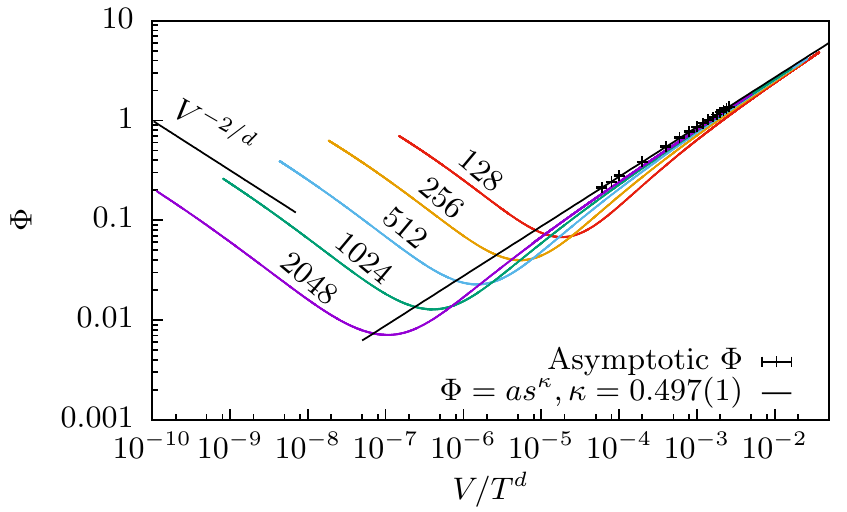}
            \caption{\label{fig:rate4D}
                (color online)
                Rate function of the distribution of the $d=4$ dimensional
                volume of the convex hull of RWs for different
                walk lengths $T$. Crosses mark the $T\to\infty$ extrapolated
                values of the asymptotic rate function as shown in
                Fig.~\ref{fig:rateExtra}. To those a power law is fitted
                yielding an estimate for the rate function consistent with the
                guess in Eq.~\eqref{eq:rateGuess}. Further, the expected
                power law behavior of the left tail is approached.
            }
        \end{figure}

        Fitting the power law Eq.~\eqref{eq:rateGuess} through the extrapolated
        points, as shown in Fig.~\ref{fig:rate4D} for the distribution of the volume in $d=4$,
        confirms the expectation of
        $\kappa = 2/\deff$. This holds also for the other cases we considered (not shown as a figure).
        All measured values of $\kappa$ are tabulated in Tab.~\ref{tab:kappa}
        and are in reasonable agreement with the expectations. Note that
        the error estimates are only statistical errors, the actual errors
        are hard to quantify, but the good $\chi^{2}_{\mathrm{red}}$ values and
        the good agreement of the exponents with the expectations, suggests
        that the power law is a reasonable ansatz and systematic errors due to
        deviations from this power law or finite-size effects are minor.

        Since the same arguments are applicable for multiple
        walks, this procedure is tested for the distributions of $m=3$ multiple
        walkers in $d=3$ dimensions, which does also yield within errorbars
        the same exponent $\kappa = 0.642(17)$ as for the single walker (no figure).

        Also note that the power-law relation for the left tail becomes visible,
        in the far left tail. The expected slope of the left tail
        $\Phi \propto s^{-2/d}$ (cf.~Eq.~\eqref{eq:leftTail}) is visualized in
        the far left tail in Fig.~\ref{fig:rate4D} and seems to be
        a reasonable approximation.

        \begin{table}[htb]
            \begin{ruledtabular}
            \begin{tabular}{r r r r r}
                \multirow{ 2}{*}{$d$} & \multicolumn{2}{c}{volume $V$}                                               & \multicolumn{2}{c}{surface $\partial V$}\\
                \cline{2-3} \cline{4-5}
                    & \multicolumn{1}{c}{expected $\kappa$} & \multicolumn{1}{c}{measured $\kappa$} & \multicolumn{1}{c}{expected $\kappa$} & \multicolumn{1}{c}{measured $\kappa$}\\[0.1cm]
                \hline
                \noalign{\vskip 0.1cm}
                $2$ & $1$   & 0.994(4) & $2$   & 1.996(2) \\
                $3$ & $2/3$ & 0.665(1) & $1$   & 0.994(2) \\
                $4$ & $1/2$ & 0.497(1) & $2/3$ & 0.647(5) \\
            \end{tabular}
            \end{ruledtabular}
            \caption{\label{tab:kappa}
                Comparison of expected and measured rate function exponent $\kappa$.
            }
        \end{table}

    \section{Conclusions}\label{sec:conclusion}
        We studied the volume and surface of convex hulls of RWs
        in up to $d=6$ dimensions for which we confirmed the analytically known
        asymptotic means and we estimated the asymptotic variances.

        Further, using sophisticated large-deviation sampling techniques we
        obtained large parts of the distributions $P$
        in up to $d=4$ dimensions down to probability densities far smaller than
        $P=10^{-1000}$.  The distributions collapse over large ranges of the support (right tail) onto
        a single curve when being rescaled with the asymptotic behavior of the means.
        For the left tail, we observe a convergence to a limiting function.
        Even more, we used our results to confirm the expected functional shapes  of the distributions in the
        left and the right tails, for finite and extrapolated values of $T$, respectively.

        We used simple arguments and numerical simulations to determine the
        scaling behavior, as well as the asymptotic behavior also for both tails of
         the rate function
        $\Phi_r(S) \propto S^{2/\deff}$ for $d \in \{3, 4\}$ and are confident
        that it is valid in arbitrary dimensions.

        For future studies, it would be interesting to investigate the properties of the convex hulls of other
        types of random walks, exhibiting non-trivial values of $\nu$, like self-avoiding walks
        or loop-erased RWs.

    \section*{Acknowledgments}
        This work was supported by the German Science Foundation (DFG) through
        the grant HA 3169/8-1.
        HS and AKH thank the LPTMS for hospitality and financial support during one and two-month
        visits, respectively, in 2016, where considerable part of the projects were performed.
        The simulations were performed at the HPC clusters HERO and CARL, both
        located at the University of Oldenburg (Germany) and funded by the DFG
        through its Major Research Instrumentation Programme
        (INST 184/108-1 FUGG and INST 184/157-1 FUGG) and the Ministry of
        Science and Culture (MWK) of the Lower Saxony State.

    \bibliography{lit}

\begin{thebibliography}{77}%
\makeatletter
\providecommand \@ifxundefined [1]{%
 \@ifx{#1\undefined}
}%
\providecommand \@ifnum [1]{%
 \ifnum #1\expandafter \@firstoftwo
 \else \expandafter \@secondoftwo
 \fi
}%
\providecommand \@ifx [1]{%
 \ifx #1\expandafter \@firstoftwo
 \else \expandafter \@secondoftwo
 \fi
}%
\providecommand \natexlab [1]{#1}%
\providecommand \enquote  [1]{``#1''}%
\providecommand \bibnamefont  [1]{#1}%
\providecommand \bibfnamefont [1]{#1}%
\providecommand \citenamefont [1]{#1}%
\providecommand \href@noop [0]{\@secondoftwo}%
\providecommand \href [0]{\begingroup \@sanitize@url \@href}%
\providecommand \@href[1]{\@@startlink{#1}\@@href}%
\providecommand \@@href[1]{\endgroup#1\@@endlink}%
\providecommand \@sanitize@url [0]{\catcode `\\12\catcode `\$12\catcode
  `\&12\catcode `\#12\catcode `\^12\catcode `\_12\catcode `\%12\relax}%
\providecommand \@@startlink[1]{}%
\providecommand \@@endlink[0]{}%
\providecommand \url  [0]{\begingroup\@sanitize@url \@url }%
\providecommand \@url [1]{\endgroup\@href {#1}{\urlprefix }}%
\providecommand \urlprefix  [0]{URL }%
\providecommand \Eprint [0]{\href }%
\providecommand \doibase [0]{http://dx.doi.org/}%
\providecommand \selectlanguage [0]{\@gobble}%
\providecommand \bibinfo  [0]{\@secondoftwo}%
\providecommand \bibfield  [0]{\@secondoftwo}%
\providecommand \translation [1]{[#1]}%
\providecommand \BibitemOpen [0]{}%
\providecommand \bibitemStop [0]{}%
\providecommand \bibitemNoStop [0]{.\EOS\space}%
\providecommand \EOS [0]{\spacefactor3000\relax}%
\providecommand \BibitemShut  [1]{\csname bibitem#1\endcsname}%
\let\auto@bib@innerbib\@empty
\bibitem [{\citenamefont {Hughes}(1996)}]{hughes1996random}%
  \BibitemOpen
  \bibfield  {author} {\bibinfo {author} {\bibfnamefont {B.~D.}\ \bibnamefont
  {Hughes}},\ }\href@noop {} {\emph {\bibinfo {title} {Random walks and random
  environments}}}\ (\bibinfo  {publisher} {Clarendon Press Oxford},\ \bibinfo
  {year} {1996})\BibitemShut {NoStop}%
\bibitem [{\citenamefont {Pearson}(1905)}]{pearson1905problem}%
  \BibitemOpen
  \bibfield  {author} {\bibinfo {author} {\bibfnamefont {K.}~\bibnamefont
  {Pearson}},\ }\href {\doibase 10.1038/072294b0} {\bibfield  {journal}
  {\bibinfo  {journal} {Nature}\ }\textbf {\bibinfo {volume} {72}},\ \bibinfo
  {pages} {294} (\bibinfo {year} {1905})}\BibitemShut {NoStop}%
\bibitem [{\citenamefont {Strutt}(1919)}]{rayleigh1919xxxi}%
  \BibitemOpen
  \bibfield  {author} {\bibinfo {author} {\bibfnamefont {J.~W.}\ \bibnamefont
  {Strutt}},\ }\href@noop {} {\bibfield  {journal} {\bibinfo  {journal} {The
  London, Edinburgh, and Dublin Philosophical Magazine and Journal of Science}\
  }\textbf {\bibinfo {volume} {37}},\ \bibinfo {pages} {321} (\bibinfo {year}
  {1919})}\BibitemShut {NoStop}%
\bibitem [{\citenamefont {P{\'o}lya}(1921)}]{polya1921}%
  \BibitemOpen
  \bibfield  {author} {\bibinfo {author} {\bibfnamefont {G.}~\bibnamefont
  {P{\'o}lya}},\ }\href {\doibase 10.1007/BF01458701} {\bibfield  {journal}
  {\bibinfo  {journal} {Mathematische Annalen}\ }\textbf {\bibinfo {volume}
  {84}},\ \bibinfo {pages} {149} (\bibinfo {year} {1921})}\BibitemShut
  {NoStop}%
\bibitem [{\citenamefont {Patlak}(1953)}]{Patlak1953random}%
  \BibitemOpen
  \bibfield  {author} {\bibinfo {author} {\bibfnamefont {C.~S.}\ \bibnamefont
  {Patlak}},\ }\href {\doibase 10.1007/BF02476407} {\bibfield  {journal}
  {\bibinfo  {journal} {The bulletin of mathematical biophysics}\ }\textbf
  {\bibinfo {volume} {15}},\ \bibinfo {pages} {311} (\bibinfo {year}
  {1953})}\BibitemShut {NoStop}%
\bibitem [{\citenamefont {Kareiva}\ and\ \citenamefont
  {Shigesada}(1983)}]{Kareiva1983analyzing}%
  \BibitemOpen
  \bibfield  {author} {\bibinfo {author} {\bibfnamefont {P.~M.}\ \bibnamefont
  {Kareiva}}\ and\ \bibinfo {author} {\bibfnamefont {N.}~\bibnamefont
  {Shigesada}},\ }\href {\doibase 10.1007/BF00379695} {\bibfield  {journal}
  {\bibinfo  {journal} {Oecologia}\ }\textbf {\bibinfo {volume} {56}},\
  \bibinfo {pages} {234} (\bibinfo {year} {1983})}\BibitemShut {NoStop}%
\bibitem [{\citenamefont {Bovet}\ and\ \citenamefont
  {Benhamou}(1988)}]{bovet1988spatial}%
  \BibitemOpen
  \bibfield  {author} {\bibinfo {author} {\bibfnamefont {P.}~\bibnamefont
  {Bovet}}\ and\ \bibinfo {author} {\bibfnamefont {S.}~\bibnamefont
  {Benhamou}},\ }\href {\doibase 10.1016/S0022-5193(88)80038-9} {\bibfield
  {journal} {\bibinfo  {journal} {Journal of Theoretical Biology}\ }\textbf
  {\bibinfo {volume} {131}},\ \bibinfo {pages} {419 } (\bibinfo {year}
  {1988})}\BibitemShut {NoStop}%
\bibitem [{\citenamefont {Madras}\ and\ \citenamefont
  {Slade}(2013)}]{Madras2013}%
  \BibitemOpen
  \bibfield  {author} {\bibinfo {author} {\bibfnamefont {N.}~\bibnamefont
  {Madras}}\ and\ \bibinfo {author} {\bibfnamefont {G.}~\bibnamefont {Slade}},\
  }\enquote {\bibinfo {title} {The self-avoiding walk},}\ \ (\bibinfo
  {publisher} {Springer New York},\ \bibinfo {address} {New York, NY},\
  \bibinfo {year} {2013})\ Chap.\ \bibinfo {chapter} {Analysis of {Monte
  Carlo}imethods}, pp.\ \bibinfo {pages} {281--364}\BibitemShut {NoStop}%
\bibitem [{\citenamefont {Lawler}(1980)}]{Lawler1980Self}%
  \BibitemOpen
  \bibfield  {author} {\bibinfo {author} {\bibfnamefont {G.~F.}\ \bibnamefont
  {Lawler}},\ }\href {\doibase 10.1215/S0012-7094-80-04741-9} {\bibfield
  {journal} {\bibinfo  {journal} {{Duke Math. J.}}\ }\textbf {\bibinfo {volume}
  {47}},\ \bibinfo {pages} {655} (\bibinfo {year} {1980})}\BibitemShut
  {NoStop}%
\bibitem [{\citenamefont {Weinrib}\ and\ \citenamefont
  {Trugman}(1985)}]{weinrib1985kinetic}%
  \BibitemOpen
  \bibfield  {author} {\bibinfo {author} {\bibfnamefont {A.}~\bibnamefont
  {Weinrib}}\ and\ \bibinfo {author} {\bibfnamefont {S.~A.}\ \bibnamefont
  {Trugman}},\ }\href {\doibase 10.1103/PhysRevB.31.2993} {\bibfield  {journal}
  {\bibinfo  {journal} {Phys. Rev. B}\ }\textbf {\bibinfo {volume} {31}},\
  \bibinfo {pages} {2993} (\bibinfo {year} {1985})}\BibitemShut {NoStop}%
\bibitem [{\citenamefont {Smoluchowski}(1916)}]{Smoluchowski1916brownsche}%
  \BibitemOpen
  \bibfield  {author} {\bibinfo {author} {\bibfnamefont {M.~V.}\ \bibnamefont
  {Smoluchowski}},\ }\href {\doibase 10.1002/andp.19163532408} {\bibfield
  {journal} {\bibinfo  {journal} {Annalen der Physik}\ }\textbf {\bibinfo
  {volume} {353}},\ \bibinfo {pages} {1103} (\bibinfo {year}
  {1916})}\BibitemShut {NoStop}%
\bibitem [{\citenamefont {Alt}(1980)}]{Alt1980biased}%
  \BibitemOpen
  \bibfield  {author} {\bibinfo {author} {\bibfnamefont {W.}~\bibnamefont
  {Alt}},\ }\href {\doibase 10.1007/BF00275919} {\bibfield  {journal} {\bibinfo
   {journal} {Journal of Mathematical Biology}\ }\textbf {\bibinfo {volume}
  {9}},\ \bibinfo {pages} {147} (\bibinfo {year} {1980})}\BibitemShut {NoStop}%
\bibitem [{\citenamefont {van Haastert}\ and\ \citenamefont
  {Postma}(2007)}]{vanHaastert2007biased}%
  \BibitemOpen
  \bibfield  {author} {\bibinfo {author} {\bibfnamefont {P.~J.}\ \bibnamefont
  {van Haastert}}\ and\ \bibinfo {author} {\bibfnamefont {M.}~\bibnamefont
  {Postma}},\ }\href {\doibase 10.1529/biophysj.107.104356} {\bibfield
  {journal} {\bibinfo  {journal} {Biophysical Journal}\ }\textbf {\bibinfo
  {volume} {93}},\ \bibinfo {pages} {1787 } (\bibinfo {year}
  {2007})}\BibitemShut {NoStop}%
\bibitem [{\citenamefont {Weesakul}(1961)}]{Weesakul1961random}%
  \BibitemOpen
  \bibfield  {author} {\bibinfo {author} {\bibfnamefont {B.}~\bibnamefont
  {Weesakul}},\ }\href {\doibase 10.2307/2237837} {\bibfield  {journal}
  {\bibinfo  {journal} {The Annals of Mathematical Statistics}\ }\textbf
  {\bibinfo {volume} {32}},\ \bibinfo {pages} {765} (\bibinfo {year}
  {1961})}\BibitemShut {NoStop}%
\bibitem [{\citenamefont {Kac}(1947)}]{kac1947random}%
  \BibitemOpen
  \bibfield  {author} {\bibinfo {author} {\bibfnamefont {M.}~\bibnamefont
  {Kac}},\ }\href {\doibase 10.2307/2304386} {\bibfield  {journal} {\bibinfo
  {journal} {The American Mathematical Monthly}\ }\textbf {\bibinfo {volume}
  {54}},\ \bibinfo {pages} {369} (\bibinfo {year} {1947})}\BibitemShut
  {NoStop}%
\bibitem [{\citenamefont {Fisher}(1984)}]{Fisher1984walks}%
  \BibitemOpen
  \bibfield  {author} {\bibinfo {author} {\bibfnamefont {M.~E.}\ \bibnamefont
  {Fisher}},\ }\href {\doibase 10.1007/BF01009436} {\bibfield  {journal}
  {\bibinfo  {journal} {Journal of Statistical Physics}\ }\textbf {\bibinfo
  {volume} {34}},\ \bibinfo {pages} {667} (\bibinfo {year} {1984})}\BibitemShut
  {NoStop}%
\bibitem [{\citenamefont {Schehr}\ \emph {et~al.}(2008)\citenamefont {Schehr},
  \citenamefont {Majumdar}, \citenamefont {Comtet},\ and\ \citenamefont
  {Randon-Furling}}]{Schehr2008Exact}%
  \BibitemOpen
  \bibfield  {author} {\bibinfo {author} {\bibfnamefont {G.}~\bibnamefont
  {Schehr}}, \bibinfo {author} {\bibfnamefont {S.~N.}\ \bibnamefont
  {Majumdar}}, \bibinfo {author} {\bibfnamefont {A.}~\bibnamefont {Comtet}}, \
  and\ \bibinfo {author} {\bibfnamefont {J.}~\bibnamefont {Randon-Furling}},\
  }\href {\doibase 10.1103/PhysRevLett.101.150601} {\bibfield  {journal}
  {\bibinfo  {journal} {Phys. Rev. Lett.}\ }\textbf {\bibinfo {volume} {101}},\
  \bibinfo {pages} {150601} (\bibinfo {year} {2008})}\BibitemShut {NoStop}%
\bibitem [{\citenamefont {Witten}\ and\ \citenamefont
  {Sander}(1983)}]{witten1983diffusion}%
  \BibitemOpen
  \bibfield  {author} {\bibinfo {author} {\bibfnamefont {T.~A.}\ \bibnamefont
  {Witten}}\ and\ \bibinfo {author} {\bibfnamefont {L.~M.}\ \bibnamefont
  {Sander}},\ }\href {\doibase 10.1103/PhysRevB.27.5686} {\bibfield  {journal}
  {\bibinfo  {journal} {Phys. Rev. B}\ }\textbf {\bibinfo {volume} {27}},\
  \bibinfo {pages} {5686} (\bibinfo {year} {1983})}\BibitemShut {NoStop}%
\bibitem [{\citenamefont {Schaefer}(1973)}]{Schaefer1973dynamics}%
  \BibitemOpen
  \bibfield  {author} {\bibinfo {author} {\bibfnamefont {D.~W.}\ \bibnamefont
  {Schaefer}},\ }\href {\doibase 10.1126/science.180.4092.1293} {\bibfield
  {journal} {\bibinfo  {journal} {Science}\ }\textbf {\bibinfo {volume}
  {180}},\ \bibinfo {pages} {1293} (\bibinfo {year} {1973})}\BibitemShut
  {NoStop}%
\bibitem [{\citenamefont {Codling}\ \emph {et~al.}(2008)\citenamefont
  {Codling}, \citenamefont {Plank},\ and\ \citenamefont
  {Benhamou}}]{Codling2008random}%
  \BibitemOpen
  \bibfield  {author} {\bibinfo {author} {\bibfnamefont {E.~A.}\ \bibnamefont
  {Codling}}, \bibinfo {author} {\bibfnamefont {M.~J.}\ \bibnamefont {Plank}},
  \ and\ \bibinfo {author} {\bibfnamefont {S.}~\bibnamefont {Benhamou}},\
  }\href {\doibase 10.1098/rsif.2008.0014} {\bibfield  {journal} {\bibinfo
  {journal} {Journal of The Royal Society Interface}\ }\textbf {\bibinfo
  {volume} {5}},\ \bibinfo {pages} {813} (\bibinfo {year} {2008})}\BibitemShut
  {NoStop}%
\bibitem [{\citenamefont {Fama}(1965)}]{Fama1965Random}%
  \BibitemOpen
  \bibfield  {author} {\bibinfo {author} {\bibfnamefont {E.~F.}\ \bibnamefont
  {Fama}},\ }\href {\doibase 10.2469/faj.v21.n5.55} {\bibfield  {journal}
  {\bibinfo  {journal} {Financial Analysts Journal}\ }\textbf {\bibinfo
  {volume} {21}},\ \bibinfo {pages} {55} (\bibinfo {year} {1965})}\BibitemShut
  {NoStop}%
\bibitem [{\citenamefont {Rosvall}\ and\ \citenamefont
  {Bergstrom}(2008)}]{Rosvall2008Maps}%
  \BibitemOpen
  \bibfield  {author} {\bibinfo {author} {\bibfnamefont {M.}~\bibnamefont
  {Rosvall}}\ and\ \bibinfo {author} {\bibfnamefont {C.~T.}\ \bibnamefont
  {Bergstrom}},\ }\href {\doibase 10.1073/pnas.0706851105} {\bibfield
  {journal} {\bibinfo  {journal} {Proceedings of the National Academy of
  Sciences}\ }\textbf {\bibinfo {volume} {105}},\ \bibinfo {pages} {1118}
  (\bibinfo {year} {2008})}\BibitemShut {NoStop}%
\bibitem [{\citenamefont {Gupta}\ \emph {et~al.}(2013)\citenamefont {Gupta},
  \citenamefont {Goel}, \citenamefont {Lin}, \citenamefont {Sharma},
  \citenamefont {Wang},\ and\ \citenamefont {Zadeh}}]{Gupta2013WTF}%
  \BibitemOpen
  \bibfield  {author} {\bibinfo {author} {\bibfnamefont {P.}~\bibnamefont
  {Gupta}}, \bibinfo {author} {\bibfnamefont {A.}~\bibnamefont {Goel}},
  \bibinfo {author} {\bibfnamefont {J.}~\bibnamefont {Lin}}, \bibinfo {author}
  {\bibfnamefont {A.}~\bibnamefont {Sharma}}, \bibinfo {author} {\bibfnamefont
  {D.}~\bibnamefont {Wang}}, \ and\ \bibinfo {author} {\bibfnamefont
  {R.}~\bibnamefont {Zadeh}},\ }in\ \href {\doibase 10.1145/2488388.2488433}
  {\emph {\bibinfo {booktitle} {Proceedings of the 22Nd International
  Conference on World Wide Web}}},\ \bibinfo {series and number} {WWW '13}\
  (\bibinfo  {publisher} {ACM},\ \bibinfo {address} {New York, NY, USA},\
  \bibinfo {year} {2013})\ pp.\ \bibinfo {pages} {505--514}\BibitemShut
  {NoStop}%
\bibitem [{\citenamefont {Dumonteil}\ \emph {et~al.}(2013)\citenamefont
  {Dumonteil}, \citenamefont {Majumdar}, \citenamefont {Rosso},\ and\
  \citenamefont {Zoia}}]{Dumonteil2013spatial}%
  \BibitemOpen
  \bibfield  {author} {\bibinfo {author} {\bibfnamefont {E.}~\bibnamefont
  {Dumonteil}}, \bibinfo {author} {\bibfnamefont {S.~N.}\ \bibnamefont
  {Majumdar}}, \bibinfo {author} {\bibfnamefont {A.}~\bibnamefont {Rosso}}, \
  and\ \bibinfo {author} {\bibfnamefont {A.}~\bibnamefont {Zoia}},\ }\href
  {\doibase 10.1073/pnas.1213237110} {\bibfield  {journal} {\bibinfo  {journal}
  {Proceedings of the National Academy of Sciences}\ }\textbf {\bibinfo
  {volume} {110}},\ \bibinfo {pages} {4239} (\bibinfo {year}
  {2013})}\BibitemShut {NoStop}%
\bibitem [{\citenamefont {Kuhn}(1934)}]{Kuhn1934gestalt}%
  \BibitemOpen
  \bibfield  {author} {\bibinfo {author} {\bibfnamefont {W.}~\bibnamefont
  {Kuhn}},\ }\href {\doibase 10.1007/BF01451681} {\bibfield  {journal}
  {\bibinfo  {journal} {Kolloid-Zeitschrift}\ }\textbf {\bibinfo {volume}
  {68}},\ \bibinfo {pages} {2} (\bibinfo {year} {1934})}\BibitemShut {NoStop}%
\bibitem [{\citenamefont {Helfand}(1975)}]{helfand1975theory}%
  \BibitemOpen
  \bibfield  {author} {\bibinfo {author} {\bibfnamefont {E.}~\bibnamefont
  {Helfand}},\ }\href {\doibase 10.1063/1.430517} {\bibfield  {journal}
  {\bibinfo  {journal} {The Journal of Chemical Physics}\ }\textbf {\bibinfo
  {volume} {62}},\ \bibinfo {pages} {999} (\bibinfo {year} {1975})}\BibitemShut
  {NoStop}%
\bibitem [{\citenamefont {Haber}\ \emph {et~al.}(2000)\citenamefont {Haber},
  \citenamefont {Ruiz},\ and\ \citenamefont {Wirtz}}]{Haber2000shape}%
  \BibitemOpen
  \bibfield  {author} {\bibinfo {author} {\bibfnamefont {C.}~\bibnamefont
  {Haber}}, \bibinfo {author} {\bibfnamefont {S.~A.}\ \bibnamefont {Ruiz}}, \
  and\ \bibinfo {author} {\bibfnamefont {D.}~\bibnamefont {Wirtz}},\ }\href
  {\doibase 10.1073/pnas.190320097} {\bibfield  {journal} {\bibinfo  {journal}
  {Proceedings of the National Academy of Sciences}\ }\textbf {\bibinfo
  {volume} {97}},\ \bibinfo {pages} {10792} (\bibinfo {year}
  {2000})}\BibitemShut {NoStop}%
\bibitem [{\citenamefont {Bartumeus}\ \emph {et~al.}(2005)\citenamefont
  {Bartumeus}, \citenamefont {da~Luz}, \citenamefont {Viswanathan},\ and\
  \citenamefont {Catalan}}]{Bartumeus2005animal}%
  \BibitemOpen
  \bibfield  {author} {\bibinfo {author} {\bibfnamefont {F.}~\bibnamefont
  {Bartumeus}}, \bibinfo {author} {\bibfnamefont {M.~G.~E.}\ \bibnamefont
  {da~Luz}}, \bibinfo {author} {\bibfnamefont {G.~M.}\ \bibnamefont
  {Viswanathan}}, \ and\ \bibinfo {author} {\bibfnamefont {J.}~\bibnamefont
  {Catalan}},\ }\href {\doibase 10.1890/04-1806} {\bibfield  {journal}
  {\bibinfo  {journal} {Ecology}\ }\textbf {\bibinfo {volume} {86}},\ \bibinfo
  {pages} {3078} (\bibinfo {year} {2005})}\BibitemShut {NoStop}%
\bibitem [{\citenamefont {B{\"o}rger}\ \emph {et~al.}(2008)\citenamefont
  {B{\"o}rger}, \citenamefont {Dalziel},\ and\ \citenamefont
  {Fryxell}}]{Borger2008general}%
  \BibitemOpen
  \bibfield  {author} {\bibinfo {author} {\bibfnamefont {L.}~\bibnamefont
  {B{\"o}rger}}, \bibinfo {author} {\bibfnamefont {B.~D.}\ \bibnamefont
  {Dalziel}}, \ and\ \bibinfo {author} {\bibfnamefont {J.~M.}\ \bibnamefont
  {Fryxell}},\ }\href {\doibase 10.1111/j.1461-0248.2008.01182.x} {\bibfield
  {journal} {\bibinfo  {journal} {Ecology Letters}\ }\textbf {\bibinfo {volume}
  {11}},\ \bibinfo {pages} {637} (\bibinfo {year} {2008})}\BibitemShut
  {NoStop}%
\bibitem [{\citenamefont {Worton}(1995)}]{worton1995convex}%
  \BibitemOpen
  \bibfield  {author} {\bibinfo {author} {\bibfnamefont {B.~J.}\ \bibnamefont
  {Worton}},\ }\href {\doibase 10.2307/2533254} {\bibfield  {journal} {\bibinfo
   {journal} {Biometrics}\ }\textbf {\bibinfo {volume} {51}},\ \bibinfo {pages}
  {1206} (\bibinfo {year} {1995})}\BibitemShut {NoStop}%
\bibitem [{\citenamefont {Giuggioli}\ \emph {et~al.}(2011)\citenamefont
  {Giuggioli}, \citenamefont {Potts},\ and\ \citenamefont
  {Harris}}]{Giuggioli2011animal}%
  \BibitemOpen
  \bibfield  {author} {\bibinfo {author} {\bibfnamefont {L.}~\bibnamefont
  {Giuggioli}}, \bibinfo {author} {\bibfnamefont {J.~R.}\ \bibnamefont
  {Potts}}, \ and\ \bibinfo {author} {\bibfnamefont {S.}~\bibnamefont
  {Harris}},\ }\href {\doibase 10.1371/journal.pcbi.1002008} {\bibfield
  {journal} {\bibinfo  {journal} {PLOS Computational Biology}\ }\textbf
  {\bibinfo {volume} {7}},\ \bibinfo {pages} {1} (\bibinfo {year}
  {2011})}\BibitemShut {NoStop}%
\bibitem [{\citenamefont {Lanoisel\'ee}\ and\ \citenamefont
  {Grebenkov}(2017)}]{grebenkov2017unraveling}%
  \BibitemOpen
  \bibfield  {author} {\bibinfo {author} {\bibfnamefont {Y.}~\bibnamefont
  {Lanoisel\'ee}}\ and\ \bibinfo {author} {\bibfnamefont {D.~S.}\ \bibnamefont
  {Grebenkov}},\ }\href {\doibase 10.1103/PhysRevE.96.022144} {\bibfield
  {journal} {\bibinfo  {journal} {Phys. Rev. E}\ }\textbf {\bibinfo {volume}
  {96}},\ \bibinfo {pages} {022144} (\bibinfo {year} {2017})}\BibitemShut
  {NoStop}%
\bibitem [{\citenamefont {G{\'e}rard~Letac}(1980)}]{Letac1980Expected}%
  \BibitemOpen
  \bibfield  {author} {\bibinfo {author} {\bibfnamefont {L.~T.}\ \bibnamefont
  {G{\'e}rard~Letac}},\ }\href {\doibase 10.2307/2322010} {\bibfield  {journal}
  {\bibinfo  {journal} {The American Mathematical Monthly}\ }\textbf {\bibinfo
  {volume} {87}},\ \bibinfo {pages} {142} (\bibinfo {year} {1980})}\BibitemShut
  {NoStop}%
\bibitem [{\citenamefont {Letac}(1993)}]{Letac1993explicit}%
  \BibitemOpen
  \bibfield  {author} {\bibinfo {author} {\bibfnamefont {G.}~\bibnamefont
  {Letac}},\ }\href {\doibase 10.1007/BF01047580} {\bibfield  {journal}
  {\bibinfo  {journal} {Journal of Theoretical Probability}\ }\textbf {\bibinfo
  {volume} {6}},\ \bibinfo {pages} {385} (\bibinfo {year} {1993})}\BibitemShut
  {NoStop}%
\bibitem [{\citenamefont {Randon-Furling}\ \emph {et~al.}(2009)\citenamefont
  {Randon-Furling}, \citenamefont {Majumdar},\ and\ \citenamefont
  {Comtet}}]{Majumdar2009Convex}%
  \BibitemOpen
  \bibfield  {author} {\bibinfo {author} {\bibfnamefont {J.}~\bibnamefont
  {Randon-Furling}}, \bibinfo {author} {\bibfnamefont {S.~N.}\ \bibnamefont
  {Majumdar}}, \ and\ \bibinfo {author} {\bibfnamefont {A.}~\bibnamefont
  {Comtet}},\ }\href {\doibase 10.1103/PhysRevLett.103.140602} {\bibfield
  {journal} {\bibinfo  {journal} {Phys. Rev. Lett.}\ }\textbf {\bibinfo
  {volume} {103}},\ \bibinfo {pages} {140602} (\bibinfo {year}
  {2009})}\BibitemShut {NoStop}%
\bibitem [{\citenamefont {Majumdar}\ \emph {et~al.}(2010)\citenamefont
  {Majumdar}, \citenamefont {Comtet},\ and\ \citenamefont
  {Randon-Furling}}]{Majumdar2010Random}%
  \BibitemOpen
  \bibfield  {author} {\bibinfo {author} {\bibfnamefont {S.~N.}\ \bibnamefont
  {Majumdar}}, \bibinfo {author} {\bibfnamefont {A.}~\bibnamefont {Comtet}}, \
  and\ \bibinfo {author} {\bibfnamefont {J.}~\bibnamefont {Randon-Furling}},\
  }\href {\doibase 10.1007/s10955-009-9905-z} {\bibfield  {journal} {\bibinfo
  {journal} {Journal of Statistical Physics}\ }\textbf {\bibinfo {volume}
  {138}},\ \bibinfo {pages} {955} (\bibinfo {year} {2010})}\BibitemShut
  {NoStop}%
\bibitem [{\citenamefont {Chupeau}\ \emph {et~al.}(2015)\citenamefont
  {Chupeau}, \citenamefont {B{\'e}nichou},\ and\ \citenamefont
  {Majumdar}}]{Chupeau2015Convex}%
  \BibitemOpen
  \bibfield  {author} {\bibinfo {author} {\bibfnamefont {M.}~\bibnamefont
  {Chupeau}}, \bibinfo {author} {\bibfnamefont {O.}~\bibnamefont
  {B{\'e}nichou}}, \ and\ \bibinfo {author} {\bibfnamefont {S.~N.}\
  \bibnamefont {Majumdar}},\ }\href {\doibase 10.1103/PhysRevE.91.050104}
  {\bibfield  {journal} {\bibinfo  {journal} {Phys. Rev. E}\ }\textbf {\bibinfo
  {volume} {91}},\ \bibinfo {pages} {050104} (\bibinfo {year}
  {2015})}\BibitemShut {NoStop}%
\bibitem [{\citenamefont {Eldan}(2014)}]{Eldan2014Volumetric}%
  \BibitemOpen
  \bibfield  {author} {\bibinfo {author} {\bibfnamefont {R.}~\bibnamefont
  {Eldan}},\ }\href {\doibase 10.1214/EJP.v19-2571} {\bibfield  {journal}
  {\bibinfo  {journal} {Electron. J. Probab.}\ }\textbf {\bibinfo {volume}
  {19}},\ \bibinfo {pages} {no. 45, 1} (\bibinfo {year} {2014})}\BibitemShut
  {NoStop}%
\bibitem [{\citenamefont {Kabluchko}\ and\ \citenamefont
  {Zaporozhets}(2016)}]{kabluchko2016intrinsic}%
  \BibitemOpen
  \bibfield  {author} {\bibinfo {author} {\bibfnamefont {Z.}~\bibnamefont
  {Kabluchko}}\ and\ \bibinfo {author} {\bibfnamefont {D.}~\bibnamefont
  {Zaporozhets}},\ }\href {\doibase 10.1090/tran/6628} {\bibfield  {journal}
  {\bibinfo  {journal} {Transactions of the American Mathematical Society}\
  }\textbf {\bibinfo {volume} {368}},\ \bibinfo {pages} {8873} (\bibinfo {year}
  {2016})}\BibitemShut {NoStop}%
\bibitem [{\citenamefont {Vysotsky}\ and\ \citenamefont
  {Zaporozhets}(2015)}]{vysotsky2015convex}%
  \BibitemOpen
  \bibfield  {author} {\bibinfo {author} {\bibfnamefont {V.}~\bibnamefont
  {Vysotsky}}\ and\ \bibinfo {author} {\bibfnamefont {D.}~\bibnamefont
  {Zaporozhets}},\ }\href@noop {} {\bibfield  {journal} {\bibinfo  {journal}
  {arXiv preprint arXiv:1506.07827}\ } (\bibinfo {year} {2015})}\BibitemShut
  {NoStop}%
\bibitem [{\citenamefont {Grebenkov}\ \emph {et~al.}(2017)\citenamefont
  {Grebenkov}, \citenamefont {Lanoisel{\'e}e},\ and\ \citenamefont
  {Majumdar}}]{grebenkov2017mean}%
  \BibitemOpen
  \bibfield  {author} {\bibinfo {author} {\bibfnamefont {D.~S.}\ \bibnamefont
  {Grebenkov}}, \bibinfo {author} {\bibfnamefont {Y.}~\bibnamefont
  {Lanoisel{\'e}e}}, \ and\ \bibinfo {author} {\bibfnamefont {S.~N.}\
  \bibnamefont {Majumdar}},\ }\href@noop {} {\bibfield  {journal} {\bibinfo
  {journal} {arXiv preprint arXiv:1706.08052}\ } (\bibinfo {year}
  {2017})}\BibitemShut {NoStop}%
\bibitem [{\citenamefont {Kampf}\ \emph {et~al.}(2012)\citenamefont {Kampf},
  \citenamefont {Last},\ and\ \citenamefont {Molchanov}}]{kampf2012convex}%
  \BibitemOpen
  \bibfield  {author} {\bibinfo {author} {\bibfnamefont {J.}~\bibnamefont
  {Kampf}}, \bibinfo {author} {\bibfnamefont {G.}~\bibnamefont {Last}}, \ and\
  \bibinfo {author} {\bibfnamefont {I.}~\bibnamefont {Molchanov}},\ }\href
  {\doibase 10.1090/S0002-9939-2012-11128-1} {\bibfield  {journal} {\bibinfo
  {journal} {Proceedings of the American Mathematical Society}\ }\textbf
  {\bibinfo {volume} {140}},\ \bibinfo {pages} {2527} (\bibinfo {year}
  {2012})}\BibitemShut {NoStop}%
\bibitem [{\citenamefont {Lukovi{\'c}}\ \emph {et~al.}(2013)\citenamefont
  {Lukovi{\'c}}, \citenamefont {Geisel},\ and\ \citenamefont
  {Eule}}]{lukovic2013area}%
  \BibitemOpen
  \bibfield  {author} {\bibinfo {author} {\bibfnamefont {M.}~\bibnamefont
  {Lukovi{\'c}}}, \bibinfo {author} {\bibfnamefont {T.}~\bibnamefont {Geisel}},
  \ and\ \bibinfo {author} {\bibfnamefont {S.}~\bibnamefont {Eule}},\ }\href
  {\doibase 10.1088/1367-2630/15/6/063034} {\bibfield  {journal} {\bibinfo
  {journal} {New Journal of Physics}\ }\textbf {\bibinfo {volume} {15}},\
  \bibinfo {pages} {063034} (\bibinfo {year} {2013})}\BibitemShut {NoStop}%
\bibitem [{\citenamefont {Reymbaut}\ \emph {et~al.}(2011)\citenamefont
  {Reymbaut}, \citenamefont {Majumdar},\ and\ \citenamefont
  {Rosso}}]{reymbaut2011convex}%
  \BibitemOpen
  \bibfield  {author} {\bibinfo {author} {\bibfnamefont {A.}~\bibnamefont
  {Reymbaut}}, \bibinfo {author} {\bibfnamefont {S.~N.}\ \bibnamefont
  {Majumdar}}, \ and\ \bibinfo {author} {\bibfnamefont {A.}~\bibnamefont
  {Rosso}},\ }\href {\doibase 10.1088/1751-8113/44/41/415001} {\bibfield
  {journal} {\bibinfo  {journal} {Journal of Physics A: Mathematical and
  Theoretical}\ }\textbf {\bibinfo {volume} {44}},\ \bibinfo {pages} {415001}
  (\bibinfo {year} {2011})}\BibitemShut {NoStop}%
\bibitem [{\citenamefont {Snyder}\ and\ \citenamefont
  {Steele}(1993)}]{snyder1993convex}%
  \BibitemOpen
  \bibfield  {author} {\bibinfo {author} {\bibfnamefont {T.~L.}\ \bibnamefont
  {Snyder}}\ and\ \bibinfo {author} {\bibfnamefont {J.~M.}\ \bibnamefont
  {Steele}},\ }\href {\doibase 10.1090/S0002-9939-1993-1169048-2} {\bibfield
  {journal} {\bibinfo  {journal} {Proceedings of the American Mathematical
  Society}\ }\textbf {\bibinfo {volume} {117}},\ \bibinfo {pages} {1165}
  (\bibinfo {year} {1993})}\BibitemShut {NoStop}%
\bibitem [{\citenamefont {Goldman}(1996{\natexlab{a}})}]{Goldman1996}%
  \BibitemOpen
  \bibfield  {author} {\bibinfo {author} {\bibfnamefont {A.}~\bibnamefont
  {Goldman}},\ }\href {\doibase 10.1007/BF01192071} {\bibfield  {journal}
  {\bibinfo  {journal} {Probability Theory and Related Fields}\ }\textbf
  {\bibinfo {volume} {105}},\ \bibinfo {pages} {57} (\bibinfo {year}
  {1996}{\natexlab{a}})}\BibitemShut {NoStop}%
\bibitem [{\citenamefont {Claussen}\ \emph {et~al.}(2015)\citenamefont
  {Claussen}, \citenamefont {Hartmann},\ and\ \citenamefont
  {Majumdar}}]{Claussen2015Convex}%
  \BibitemOpen
  \bibfield  {author} {\bibinfo {author} {\bibfnamefont {G.}~\bibnamefont
  {Claussen}}, \bibinfo {author} {\bibfnamefont {A.~K.}\ \bibnamefont
  {Hartmann}}, \ and\ \bibinfo {author} {\bibfnamefont {S.~N.}\ \bibnamefont
  {Majumdar}},\ }\href {\doibase 10.1103/PhysRevE.91.052104} {\bibfield
  {journal} {\bibinfo  {journal} {Phys. Rev. E}\ }\textbf {\bibinfo {volume}
  {91}},\ \bibinfo {pages} {052104} (\bibinfo {year} {2015})}\BibitemShut
  {NoStop}%
\bibitem [{\citenamefont {Dewenter}\ \emph {et~al.}(2016)\citenamefont
  {Dewenter}, \citenamefont {Claussen}, \citenamefont {Hartmann},\ and\
  \citenamefont {Majumdar}}]{Dewenter2016Convex}%
  \BibitemOpen
  \bibfield  {author} {\bibinfo {author} {\bibfnamefont {T.}~\bibnamefont
  {Dewenter}}, \bibinfo {author} {\bibfnamefont {G.}~\bibnamefont {Claussen}},
  \bibinfo {author} {\bibfnamefont {A.~K.}\ \bibnamefont {Hartmann}}, \ and\
  \bibinfo {author} {\bibfnamefont {S.~N.}\ \bibnamefont {Majumdar}},\ }\href
  {\doibase 10.1103/PhysRevE.94.052120} {\bibfield  {journal} {\bibinfo
  {journal} {Phys. Rev. E}\ }\textbf {\bibinfo {volume} {94}},\ \bibinfo
  {pages} {052120} (\bibinfo {year} {2016})}\BibitemShut {NoStop}%
\bibitem [{\citenamefont {Akopyan}\ and\ \citenamefont
  {Vysotsky}(2016)}]{akopyan2016large}%
  \BibitemOpen
  \bibfield  {author} {\bibinfo {author} {\bibfnamefont {A.}~\bibnamefont
  {Akopyan}}\ and\ \bibinfo {author} {\bibfnamefont {V.}~\bibnamefont
  {Vysotsky}},\ }\href@noop {} {\bibfield  {journal} {\bibinfo  {journal}
  {arXiv preprint arXiv:1606.07141}\ } (\bibinfo {year} {2016})}\BibitemShut
  {NoStop}%
\bibitem [{\citenamefont {Duda}\ \emph {et~al.}(2012)\citenamefont {Duda},
  \citenamefont {Hart},\ and\ \citenamefont {Stork}}]{duda2012pattern}%
  \BibitemOpen
  \bibfield  {author} {\bibinfo {author} {\bibfnamefont {R.}~\bibnamefont
  {Duda}}, \bibinfo {author} {\bibfnamefont {P.}~\bibnamefont {Hart}}, \ and\
  \bibinfo {author} {\bibfnamefont {D.}~\bibnamefont {Stork}},\ }\href
  {https://books.google.de/books?id=Br33IRC3PkQC} {\emph {\bibinfo {title}
  {Pattern Classification}}}\ (\bibinfo  {publisher} {Wiley},\ \bibinfo {year}
  {2012})\BibitemShut {NoStop}%
\bibitem [{\citenamefont {Cornwell}\ \emph {et~al.}(2006)\citenamefont
  {Cornwell}, \citenamefont {Schwilk},\ and\ \citenamefont
  {Ackerly}}]{Cornwell2006Trait}%
  \BibitemOpen
  \bibfield  {author} {\bibinfo {author} {\bibfnamefont {W.~K.}\ \bibnamefont
  {Cornwell}}, \bibinfo {author} {\bibfnamefont {D.~W.}\ \bibnamefont
  {Schwilk}}, \ and\ \bibinfo {author} {\bibfnamefont {D.~D.}\ \bibnamefont
  {Ackerly}},\ }\href {\doibase 10.1890/0012-9658(2006)87[1465:ATTFHF]2.0.CO;2}
  {\bibfield  {journal} {\bibinfo  {journal} {Ecology}\ }\textbf {\bibinfo
  {volume} {87}},\ \bibinfo {pages} {1465} (\bibinfo {year}
  {2006})}\BibitemShut {NoStop}%
\bibitem [{\citenamefont {Preparata}\ and\ \citenamefont
  {Shamos}(1985)}]{Preparata1985convex}%
  \BibitemOpen
  \bibfield  {author} {\bibinfo {author} {\bibfnamefont {F.~P.}\ \bibnamefont
  {Preparata}}\ and\ \bibinfo {author} {\bibfnamefont {M.~I.}\ \bibnamefont
  {Shamos}},\ }\enquote {\bibinfo {title} {Convex hulls: Basic algorithms},}\
  in\ \href {\doibase 10.1007/978-1-4612-1098-6_3} {\emph {\bibinfo {booktitle}
  {Computational Geometry: An Introduction}}}\ (\bibinfo  {publisher} {Springer
  New York},\ \bibinfo {address} {New York, NY},\ \bibinfo {year} {1985})\ pp.\
  \bibinfo {pages} {95--149}\BibitemShut {NoStop}%
\bibitem [{\citenamefont {Jayaram}\ and\ \citenamefont
  {Fleyeh}(2016)}]{jayaram2016convex}%
  \BibitemOpen
  \bibfield  {author} {\bibinfo {author} {\bibfnamefont {M.}~\bibnamefont
  {Jayaram}}\ and\ \bibinfo {author} {\bibfnamefont {H.}~\bibnamefont
  {Fleyeh}},\ }\href {\doibase 10.5923/j.ajis.20160602.03} {\bibfield
  {journal} {\bibinfo  {journal} {American Journal of Intelligent Systems}\
  }\textbf {\bibinfo {volume} {6}},\ \bibinfo {pages} {48} (\bibinfo {year}
  {2016})}\BibitemShut {NoStop}%
\bibitem [{\citenamefont {Brown}(1979)}]{brown1979Voronoi}%
  \BibitemOpen
  \bibfield  {author} {\bibinfo {author} {\bibfnamefont {K.~Q.}\ \bibnamefont
  {Brown}},\ }\href {\doibase 10.1016/0020-0190(79)90074-7} {\bibfield
  {journal} {\bibinfo  {journal} {Information Processing Letters}\ }\textbf
  {\bibinfo {volume} {9}},\ \bibinfo {pages} {223 } (\bibinfo {year}
  {1979})}\BibitemShut {NoStop}%
\bibitem [{\citenamefont {Aurenhammer}(1991)}]{Aurenhammer1991Voronoi}%
  \BibitemOpen
  \bibfield  {author} {\bibinfo {author} {\bibfnamefont {F.}~\bibnamefont
  {Aurenhammer}},\ }\href {\doibase 10.1145/116873.116880} {\bibfield
  {journal} {\bibinfo  {journal} {ACM Comput. Surv.}\ }\textbf {\bibinfo
  {volume} {23}},\ \bibinfo {pages} {345} (\bibinfo {year} {1991})}\BibitemShut
  {NoStop}%
\bibitem [{\citenamefont {Klee}(1980)}]{klee1980complexity}%
  \BibitemOpen
  \bibfield  {author} {\bibinfo {author} {\bibfnamefont {V.}~\bibnamefont
  {Klee}},\ }\href {\doibase 10.1007/BF01224932} {\bibfield  {journal}
  {\bibinfo  {journal} {Archiv der Mathematik}\ }\textbf {\bibinfo {volume}
  {34}},\ \bibinfo {pages} {75} (\bibinfo {year} {1980})}\BibitemShut {NoStop}%
\bibitem [{\citenamefont {Seidel}(1981)}]{seidel1981convex}%
  \BibitemOpen
  \bibfield  {author} {\bibinfo {author} {\bibfnamefont {R.}~\bibnamefont
  {Seidel}},\ }\emph {\bibinfo {title} {A convex hull algorithm optimal for
  point sets in even dimensions}},\ \href {\doibase 10.14288/1.0051821} {Ph.D.
  thesis},\ \bibinfo  {school} {University of British Columbia} (\bibinfo
  {year} {1981})\BibitemShut {NoStop}%
\bibitem [{\citenamefont {Clarkson}\ and\ \citenamefont
  {Shor}(1989)}]{clarkson1989applications}%
  \BibitemOpen
  \bibfield  {author} {\bibinfo {author} {\bibfnamefont {K.~L.}\ \bibnamefont
  {Clarkson}}\ and\ \bibinfo {author} {\bibfnamefont {P.~W.}\ \bibnamefont
  {Shor}},\ }\href {\doibase 10.1007/BF02187740} {\bibfield  {journal}
  {\bibinfo  {journal} {Discrete {\&} Computational Geometry}\ }\textbf
  {\bibinfo {volume} {4}},\ \bibinfo {pages} {387} (\bibinfo {year}
  {1989})}\BibitemShut {NoStop}%
\bibitem [{\citenamefont {Xu}\ \emph {et~al.}(1998)\citenamefont {Xu},
  \citenamefont {Zhang},\ and\ \citenamefont {Leung}}]{Xu1998approximate}%
  \BibitemOpen
  \bibfield  {author} {\bibinfo {author} {\bibfnamefont {Z.-B.}\ \bibnamefont
  {Xu}}, \bibinfo {author} {\bibfnamefont {J.-S.}\ \bibnamefont {Zhang}}, \
  and\ \bibinfo {author} {\bibfnamefont {Y.-W.}\ \bibnamefont {Leung}},\ }\href
  {\doibase 10.1016/S0096-3003(97)10043-1} {\bibfield  {journal} {\bibinfo
  {journal} {Applied Mathematics and Computation}\ }\textbf {\bibinfo {volume}
  {94}},\ \bibinfo {pages} {193 } (\bibinfo {year} {1998})}\BibitemShut
  {NoStop}%
\bibitem [{\citenamefont {Sartipizadeh}\ and\ \citenamefont
  {Vincent}(2016)}]{Sartipizadeh2016computing}%
  \BibitemOpen
  \bibfield  {author} {\bibinfo {author} {\bibfnamefont {H.}~\bibnamefont
  {Sartipizadeh}}\ and\ \bibinfo {author} {\bibfnamefont {T.~L.}\ \bibnamefont
  {Vincent}},\ }\href@noop {} {\enquote {\bibinfo {title} {Computing the
  approximate convex hull in high dimensions},}\ } (\bibinfo {year}
  {2016})\BibitemShut {NoStop}%
\bibitem [{\citenamefont {Eddy}(1977)}]{Eddy1977Convex}%
  \BibitemOpen
  \bibfield  {author} {\bibinfo {author} {\bibfnamefont {W.~F.}\ \bibnamefont
  {Eddy}},\ }\href {\doibase 10.1145/355759.355766} {\bibfield  {journal}
  {\bibinfo  {journal} {ACM Trans. Math. Softw.}\ }\textbf {\bibinfo {volume}
  {3}},\ \bibinfo {pages} {398} (\bibinfo {year} {1977})}\BibitemShut {NoStop}%
\bibitem [{\citenamefont {Bykat}(1978)}]{Bykat1978Convex}%
  \BibitemOpen
  \bibfield  {author} {\bibinfo {author} {\bibfnamefont {A.}~\bibnamefont
  {Bykat}},\ }\href {\doibase 10.1016/0020-0190(78)90021-2} {\bibfield
  {journal} {\bibinfo  {journal} {Information Processing Letters}\ }\textbf
  {\bibinfo {volume} {7}},\ \bibinfo {pages} {296 } (\bibinfo {year}
  {1978})}\BibitemShut {NoStop}%
\bibitem [{\citenamefont {M{\"u}cke}(2009)}]{Mucke2009Quickhull}%
  \BibitemOpen
  \bibfield  {author} {\bibinfo {author} {\bibfnamefont {E.}~\bibnamefont
  {M{\"u}cke}},\ }\href {\doibase 10.1109/MCSE.2009.136} {\bibfield  {journal}
  {\bibinfo  {journal} {Computing in Science {\&} Engineering}\ }\textbf
  {\bibinfo {volume} {11}},\ \bibinfo {pages} {54} (\bibinfo {year}
  {2009})}\BibitemShut {NoStop}%
\bibitem [{\citenamefont {Barber}\ \emph {et~al.}(1996)\citenamefont {Barber},
  \citenamefont {Dobkin},\ and\ \citenamefont
  {Huhdanpaa}}]{Barber1996thequickhull}%
  \BibitemOpen
  \bibfield  {author} {\bibinfo {author} {\bibfnamefont {C.~B.}\ \bibnamefont
  {Barber}}, \bibinfo {author} {\bibfnamefont {D.~P.}\ \bibnamefont {Dobkin}},
  \ and\ \bibinfo {author} {\bibfnamefont {H.}~\bibnamefont {Huhdanpaa}},\
  }\href {\doibase 10.1.1.117.405} {\bibfield  {journal} {\bibinfo  {journal}
  {ACM Trans. Math. Softw.}\ }\textbf {\bibinfo {volume} {22}},\ \bibinfo
  {pages} {469} (\bibinfo {year} {1996})}\BibitemShut {NoStop}%
\bibitem [{\citenamefont {Wang}\ and\ \citenamefont
  {Landau}(2001{\natexlab{a}})}]{Wang2001Efficient}%
  \BibitemOpen
  \bibfield  {author} {\bibinfo {author} {\bibfnamefont {F.}~\bibnamefont
  {Wang}}\ and\ \bibinfo {author} {\bibfnamefont {D.~P.}\ \bibnamefont
  {Landau}},\ }\href {\doibase 10.1103/PhysRevLett.86.2050} {\bibfield
  {journal} {\bibinfo  {journal} {Phys. Rev. Lett.}\ }\textbf {\bibinfo
  {volume} {86}},\ \bibinfo {pages} {2050} (\bibinfo {year}
  {2001}{\natexlab{a}})}\BibitemShut {NoStop}%
\bibitem [{\citenamefont {Wang}\ and\ \citenamefont
  {Landau}(2001{\natexlab{b}})}]{Wang2001Determining}%
  \BibitemOpen
  \bibfield  {author} {\bibinfo {author} {\bibfnamefont {F.}~\bibnamefont
  {Wang}}\ and\ \bibinfo {author} {\bibfnamefont {D.~P.}\ \bibnamefont
  {Landau}},\ }\href {\doibase 10.1103/PhysRevE.64.056101} {\bibfield
  {journal} {\bibinfo  {journal} {Phys. Rev. E}\ }\textbf {\bibinfo {volume}
  {64}},\ \bibinfo {pages} {056101} (\bibinfo {year}
  {2001}{\natexlab{b}})}\BibitemShut {NoStop}%
\bibitem [{\citenamefont {Schulz}\ \emph {et~al.}(2003)\citenamefont {Schulz},
  \citenamefont {Binder}, \citenamefont {M\"uller},\ and\ \citenamefont
  {Landau}}]{Schulz2003Avoiding}%
  \BibitemOpen
  \bibfield  {author} {\bibinfo {author} {\bibfnamefont {B.~J.}\ \bibnamefont
  {Schulz}}, \bibinfo {author} {\bibfnamefont {K.}~\bibnamefont {Binder}},
  \bibinfo {author} {\bibfnamefont {M.}~\bibnamefont {M\"uller}}, \ and\
  \bibinfo {author} {\bibfnamefont {D.~P.}\ \bibnamefont {Landau}},\ }\href
  {\doibase 10.1103/PhysRevE.67.067102} {\bibfield  {journal} {\bibinfo
  {journal} {Phys. Rev. E}\ }\textbf {\bibinfo {volume} {67}},\ \bibinfo
  {pages} {067102} (\bibinfo {year} {2003})}\BibitemShut {NoStop}%
\bibitem [{\citenamefont {Belardinelli}\ and\ \citenamefont
  {Pereyra}(2007{\natexlab{a}})}]{Belardinelli2007Fast}%
  \BibitemOpen
  \bibfield  {author} {\bibinfo {author} {\bibfnamefont {R.~E.}\ \bibnamefont
  {Belardinelli}}\ and\ \bibinfo {author} {\bibfnamefont {V.~D.}\ \bibnamefont
  {Pereyra}},\ }\href {\doibase 10.1103/PhysRevE.75.046701} {\bibfield
  {journal} {\bibinfo  {journal} {Phys. Rev. E}\ }\textbf {\bibinfo {volume}
  {75}},\ \bibinfo {pages} {046701} (\bibinfo {year}
  {2007}{\natexlab{a}})}\BibitemShut {NoStop}%
\bibitem [{\citenamefont {Belardinelli}\ and\ \citenamefont
  {Pereyra}(2007{\natexlab{b}})}]{Belardinelli2007theoretical}%
  \BibitemOpen
  \bibfield  {author} {\bibinfo {author} {\bibfnamefont {R.~E.}\ \bibnamefont
  {Belardinelli}}\ and\ \bibinfo {author} {\bibfnamefont {V.~D.}\ \bibnamefont
  {Pereyra}},\ }\href {\doibase 10.1063/1.2803061} {\bibfield  {journal}
  {\bibinfo  {journal} {The Journal of Chemical Physics}\ }\textbf {\bibinfo
  {volume} {127}},\ \bibinfo {eid} {184105} (\bibinfo {year}
  {2007}{\natexlab{b}}),\ 10.1063/1.2803061}\BibitemShut {NoStop}%
\bibitem [{\citenamefont {Lee}(1993)}]{Lee1993Entropic}%
  \BibitemOpen
  \bibfield  {author} {\bibinfo {author} {\bibfnamefont {J.}~\bibnamefont
  {Lee}},\ }\href {\doibase 10.1103/PhysRevLett.71.211} {\bibfield  {journal}
  {\bibinfo  {journal} {Phys. Rev. Lett.}\ }\textbf {\bibinfo {volume} {71}},\
  \bibinfo {pages} {211} (\bibinfo {year} {1993})}\BibitemShut {NoStop}%
\bibitem [{\citenamefont {Dickman}\ and\ \citenamefont
  {Cunha-Netto}(2011)}]{Dickman2011Complete}%
  \BibitemOpen
  \bibfield  {author} {\bibinfo {author} {\bibfnamefont {R.}~\bibnamefont
  {Dickman}}\ and\ \bibinfo {author} {\bibfnamefont {A.~G.}\ \bibnamefont
  {Cunha-Netto}},\ }\href {\doibase 10.1103/PhysRevE.84.026701} {\bibfield
  {journal} {\bibinfo  {journal} {Phys. Rev. E}\ }\textbf {\bibinfo {volume}
  {84}},\ \bibinfo {pages} {026701} (\bibinfo {year} {2011})}\BibitemShut
  {NoStop}%
\bibitem [{\citenamefont {Vogel}\ \emph {et~al.}(2013)\citenamefont {Vogel},
  \citenamefont {Li}, \citenamefont {W\"ust},\ and\ \citenamefont
  {Landau}}]{landau2013generic}%
  \BibitemOpen
  \bibfield  {author} {\bibinfo {author} {\bibfnamefont {T.}~\bibnamefont
  {Vogel}}, \bibinfo {author} {\bibfnamefont {Y.~W.}\ \bibnamefont {Li}},
  \bibinfo {author} {\bibfnamefont {T.}~\bibnamefont {W\"ust}}, \ and\ \bibinfo
  {author} {\bibfnamefont {D.~P.}\ \bibnamefont {Landau}},\ }\href {\doibase
  10.1103/PhysRevLett.110.210603} {\bibfield  {journal} {\bibinfo  {journal}
  {Phys. Rev. Lett.}\ }\textbf {\bibinfo {volume} {110}},\ \bibinfo {pages}
  {210603} (\bibinfo {year} {2013})}\BibitemShut {NoStop}%
\bibitem [{\citenamefont {Goldman}(1996{\natexlab{b}})}]{Goldman1996spectrum}%
  \BibitemOpen
  \bibfield  {author} {\bibinfo {author} {\bibfnamefont {A.}~\bibnamefont
  {Goldman}},\ }\href {\doibase 10.1007/BF01192071} {\bibfield  {journal}
  {\bibinfo  {journal} {J. Prob. Theor. Relat. Fields}\ }\textbf {\bibinfo
  {volume} {105}},\ \bibinfo {pages} {57} (\bibinfo {year}
  {1996}{\natexlab{b}})}\BibitemShut {NoStop}%
\bibitem [{\citenamefont {Wade}\ and\ \citenamefont
  {Xu}(2015)}]{wade2015convex}%
  \BibitemOpen
  \bibfield  {author} {\bibinfo {author} {\bibfnamefont {A.~R.}\ \bibnamefont
  {Wade}}\ and\ \bibinfo {author} {\bibfnamefont {C.}~\bibnamefont {Xu}},\
  }\href {\doibase http://dx.doi.org/10.1016/j.spa.2015.06.008} {\bibfield
  {journal} {\bibinfo  {journal} {Stochastic Processes and their Applications}\
  }\textbf {\bibinfo {volume} {125}},\ \bibinfo {pages} {4300 } (\bibinfo
  {year} {2015})}\BibitemShut {NoStop}%
\bibitem [{\citenamefont {Engel}(2016)}]{engel2016disputation}%
  \BibitemOpen
  \bibfield  {author} {\bibinfo {author} {\bibfnamefont {A.}~\bibnamefont
  {Engel}},\ }\href@noop {} {} (\bibinfo {year} {2016}),\ \bibinfo {note}
  {personal communications}\BibitemShut {NoStop}%
\bibitem [{\citenamefont {Kundu}\ \emph {et~al.}(2013)\citenamefont {Kundu},
  \citenamefont {Majumdar},\ and\ \citenamefont {Schehr}}]{Kundu2013exact}%
  \BibitemOpen
  \bibfield  {author} {\bibinfo {author} {\bibfnamefont {A.}~\bibnamefont
  {Kundu}}, \bibinfo {author} {\bibfnamefont {S.~N.}\ \bibnamefont {Majumdar}},
  \ and\ \bibinfo {author} {\bibfnamefont {G.}~\bibnamefont {Schehr}},\ }\href
  {\doibase 10.1103/PhysRevLett.110.220602} {\bibfield  {journal} {\bibinfo
  {journal} {Phys. Rev. Lett.}\ }\textbf {\bibinfo {volume} {110}},\ \bibinfo
  {pages} {220602} (\bibinfo {year} {2013})}\BibitemShut {NoStop}%
\bibitem [{\citenamefont {Touchette}(2009)}]{Touchette2009large}%
  \BibitemOpen
  \bibfield  {author} {\bibinfo {author} {\bibfnamefont {H.}~\bibnamefont
  {Touchette}},\ }\href {\doibase 10.1016/j.physrep.2009.05.002} {\bibfield
  {journal} {\bibinfo  {journal} {Physics Reports}\ }\textbf {\bibinfo {volume}
  {478}},\ \bibinfo {pages} {1 } (\bibinfo {year} {2009})}\BibitemShut
  {NoStop}%
\end{thebibliography}%

\end{document}